\pdfoutput=1
\documentclass[pra,twocolumn,preprintnumbers,amsmath,aps,amssymb]{revtex4}
\usepackage[pdftex,bookmarks=true,colorlinks,linkcolor=red,urlcolor=blue,citecolor=blue]{hyperref}
\usepackage{bm}
\usepackage{graphicx,amsmath,amsfonts}
\usepackage[usenames]{color}
\newcommand{\be}{\begin{equation}}
\newcommand{\ee}{\end{equation}}
\newcommand{\beq}{\begin{eqnarray}}
\newcommand{\eeq}{\end{eqnarray}}
\newcommand{\tz}[2]{|t^z_{#1,#2}\rangle}
\newcommand{\figpreamble}{}
\usepackage{color}

\begin{document}

\title{Quantum Many-Body Dynamics of Coupled Double-Well Superlattices }

\author{Peter Barmettler$^1$, Ana Maria Rey$^2$, Eugene Demler$^3$, Mikhail D. Lukin$^3$, Immanuel Bloch$^4$, Vladimir Gritsev$^3$}
\affiliation {$^1$Department of Physics, University of Fribourg, CH-1700 Fribourg, Switzerland\\
$^2$Institute of Theoretical Atomic, Molecular and Optical Physics,
Harvard University, Cambridge, MA 02138 \\ $^3$Department of
Physics, Harvard University, Cambridge, MA 02138\\
$^4$ Johannes Gutenberg-Universit\"at, Institut f\"ur Physik,
Staudingerweg 7, 55099 Mainz, Germany }

\begin{abstract}

We propose a method for  controllable generation of
non-local entangled pairs using spinor atoms loaded in an
optical  superlattice. Our scheme iteratively increases the distance between entangled atoms  by controlling the coupling
between the double wells. When implemented in a finite linear chain
of $2N$ atoms, it creates a triplet valence bond state with large
persistency of entanglement (of the order of $N$). We also study the non-equilibrium dynamics of the one-dimensional ferromagnetic Heisenberg Hamiltonian and show that the time evolution of a state of decoupled triplets on each double well 
leads  to the formation of a  highly entangled state where
short-distance antiferromagnetic  correlations coexist with longer-distance  ferromagnetic ones. 
We present methods for detection and characterization of the various
dynamically generated states. These ideas are a step forward towards
the use of atoms trapped by light as  quantum information processors
and quantum simulators.
\end{abstract}

\maketitle

\section{Introduction}
The generation and manipulation  of 
entanglement  have been identified as important  requirements for
quantum teleportation~\cite{Bennett2}, quantum information
processing \cite{Nielsen} and quantum communication~\cite{Gisin}.
Engineering
 long-ranged entangled pairs in optical lattices can
also have fundamental implications  in the context of  quantum
magnetism. For example,  many   frustrated spin states such as spin liquid states correspond to coherent superpositions  of
spin singlet states \cite{Anderson}.

Recent experiments have made progress towards generating
multiparticle entanglement among an ensemble of atoms confined in optical lattices
by using controlled collisions between individual neighboring atoms
\cite{Bloch1}.  However, the generation of
long-distance pair entanglement in systems with short-range
interaction between particles (such as optical lattices) is not an
easy task.  In recent proposals long-distance EPR pairs~\cite{Bennett2} are
generated by first creating an entangled pair of quantum particles
in one location and then physically transporting one member of the
pair to another location \cite{Calarco}. However, decoherence
during the transport reduces the quality (fidelity) of the entanglement.

Our approach is based  on  coherent manipulations of triplet or singlet pairs of
ultra-cold atoms loaded in an array of double-well potentials called superlattice \cite{sebby-2006,lee-2007,folling-2007}.
These  manipulations, applied to isolated double wells, were used
for  the recent  observation of superexchange  interactions
in optical lattices \cite{RGBDL, Bloch4,vaucher-2007}.  Here we generalize these approaches to  study the many-body dynamics that arises when coupling between the double wells is allowed for. We propose various schemes that result in controllable generation of multiparticle entanglement. Specifically, we  first  discuss  a protocol that creates  from a system of spinor bosonic atoms initially prepared as an array of triplet (singlet) pairs on neighboring sites,
an array of long-distance triplet (singlet) pairs across
the lattice. The method consists of a simple iterative swapping procedure, performed by controlling  the double-well barrier height (see Fig.~\ref{cartoon}), which enables parallel generation of long-distance EPR pairs. 

We find that by combining the
iterative swapping procedure with the boundary effects always
present in a finite linear chain, one can engineer a state in which
each atom located in the right half of the superlattice is
entangled with an atom in the left half. This bipartition of
the system into its left and right parts exhibits maximal
entanglement entropy. Similar procedures have been proposed for coherently transporting quantum information \cite{romero-isart-2008} and for creating bosonic cooper-like pairs \cite{keilmann-2008} in optical lattices.  Additionally, we show that the  parallel generation of an array of
EPR pairs can be useful for efficient implementation of
entanglement purification schemes \cite{Bennett3}, which aim to
distill the few high-fidelity entangled pairs from the numerous low-fidelity ones.  

The swapping procedure described above is implemented in an array of
decoupled double wells. An interesting question that naturally
arises is  what happens with the state if the double wells are no longer
completely decoupled, but instead there exists a finite
tunneling between them. The resulting dynamics goes beyond the simple
two-particle physics behind the swapping procedure and the
experiments which control superexchange interactions~\cite{Bloch4}.
The emerging state is the consequence of {\it many-body dynamics} of
a global interacting Hamiltonian and does not require manipulations
on individually accessed atoms. This is a promising approach for creating new magnetic phases without explicitly processing a
quantum-computer protocol. Although we believe that the phenomena we
discuss here are very general, to be specific we consider in this
paper a one-dimensional chain and focus 
on the coherent evolution of the product state of triplets or singlets in each individual double well (Fig. 1a). These are dimerized states which break
translational symmetry. This choice of initial states is motivated by the fact they can be prepared in experiments \cite{Bloch4}.

Our analysis shows that the time evolution of the triplet product
state leads to the formation of a magnetic state with mixed
correlations and a high degree of multiparticle entanglement, where short-range antiferromagnetic and long-range ferromagnetic correlations coexist. This state can be experimentally probed
by measuring the singlet-triplet populations \cite{RGBDL} and
density-density correlations after time of flight \cite{ADL}. We
also find total (partial) restoration of the translational
(rotational) symmetry, which suggests that our final state has some
type of {\it spin liquid} character. By this we mean a state with strong intrinsic
fluctuations but no broken symmetries \cite{Anderson}, what may be
different from other definitions which are based on the topological order
of the quantum state \cite{fisher-1999}.

The time evolution of the initial singlet state also leads to the
restoration of the translational symmetry and high multiparticle entanglement but in this case we do not observe the strongly mixed correlations. The dynamic state has purely
antiferromagnetic character, although with an unusual behavior of
long-range correlations.

The paper is organized as follows: After introducing in
Sec.~\ref{form} the formalism and numerical techniques we use for
our analysis, in Sec.~\ref{ham-proc} we describe the basic
Hamiltonian and its possible implementation in the context of
recent experiments using optical superlattices. In
Sec.~\ref{single_switch} we present the  swapping procedure which we
refer to as a single switch dynamics and  in Sec.~\ref{many_switch}
we discuss the idea of iterative repetition of the switch as a means
to  generate long-distance entangled pairs. We also study possible
ways to experimentally detect such long-range correlations. In
Sec.~\ref{uniform} we relax the isolated double-well constraint and
allow for a finite coupling between the double wells. Specifically, we
concentrate our analysis on the many-body dynamics that emerges when
both the intra- and inter-well couplings are equal and study  the
coherent dynamics starting from both an initially prepared  triplet
product state and an initially prepared  singlet product state.
Finally, we present our conclusions in Sec.~\ref{concl}.

\section{The Formalism \label{form}}
The focus of this paper is twofold. On  one hand we study
experimentally relevant observables which can be used to detect and
characterize the dynamics of cold atoms. On the other hand we
analyze properties of entanglement in the system. The propagation and
redistribution of entanglement are not only important from the
quantum-information perspective, but  can also  help to understand
the quasiparticle dynamics as demonstrated recently
\cite{CC:Entanglement}. Such properties  are best discussed in terms
of  the entanglement entropy which corresponds to the von Neumann
entropy of the reduced density matrix with respect to a bipartition
into two subsystems \cite{Bennett}. The entanglement entropy is
defined as $S=-\mbox{tr}(\rho \log_2( \rho) )$, where the reduced
density matrix $\rho = \mbox{tr} |\psi(t)\rangle\langle \psi(t)|$ is
the trace over the states of either of the two subsystems. For the one-dimensional
systems with open boundary conditions, we will study the entanglement entropy $S_l$ of a
block of size $l$ located at the edge of the chain. In the case of an infinite system we define $S^{even}_{\infty}$ ($S^{odd}_{\infty}$) as the entropy of subsystems formed by partitioning the chain at an even (odd) bond. While any product state (a state
that can be represented as a tensor product of two pure subsystem states)
has zero entanglement entropy, maximally entanglement states at half
bipartition have entanglement entropy of $S=N$.

We use both numerical and analytic techniques to study the quantum dynamics.
For the numerical treatment we adopt the time-evolving block decimation algorithm (TEBD) for finite
\cite{Vidal:TEBD,daley:tdmrg} and periodic infinite systems \cite{Vidal:ITEBD},
which uses a matrix-product state representation and a Suzuki-Trotter
decomposition of the evolution operator. It retains only states with the lowest weights in the reduced density matrix, keeping the number of
states $\chi$ (the dimension of the matrices) finite. Consequently,
the wave-function of weakly entangled states can be handled efficiently, with the computation times of the order of $O(\chi^3N)$.

During the time evolution $\chi$ has to be increased in order to
reproduce the growing entanglement in the system. The accuracy of
the method is estimated by varying both $\chi$ and the
Suzuki-Trotter slicing \cite{Gobert}. For short and intermediate
times the TEBD algorithm  allows us to get very precise results, but
at the moment when the entanglement entropy exceeds $log_2(\chi)$,
the matrix-product representation becomes no longer accurate. To deal
with the evolution over long periods of time ($t\rightarrow\infty$), we use exact
diagonalization \cite{Hochbruck:Lanczos} techniques. Even though
these techniques can only deal with systems with small number of
lattice sites (up to 24 sites) and suffer from recurrence
effects, they are relevant for realistic setups in 1D
experiments~\cite{paredes}.

\section{Setup and procedures\label{ham-proc}}

\subsection{Effective Hamiltonian\label{eff_ham}}
We consider a system of $2N$ ultracold bosonic atoms with two
relevant hyperfine states, which we denote as $\uparrow$ and $\downarrow$,
confined within a double-well superlattice with the filling factor of 1.  The
latter can be experimentally implemented by superimposing two independent lattices one with twice the
period of the other \cite{folling-2007,Bloch4}.

In the deep barrier regime, the vibrational energy of each well,
$\hbar \omega_0$, is the  largest energy scale in the system  and
one can restrict the dynamics to the lowest vibrational states. When
restricted to the lowest band, there are three relevant energy
scales: the intra-well hopping amplitude $t_{in}$, the inter-well hopping amplitude
$t_{out}$ and the on-site interaction energy $U$. In the limit of
large $U\gg t_{in},t_{out}$ we are focusing on, the system is in the
Mott insulating regime and the only populated states are the singly
occupied ones. The  spin dynamics is described by the following
effective Hamiltonian, which takes into account the coupling between the
different singly occupied states by virtual particle-hole
excitations \cite{Duan-2003,kuklov-2003,Bloch4},
\beq H^{eff}=-J_{1}\sum_{j}{\bf S}_{2j}\cdot {\bf
S}_{2j+1}-J_{2}\sum_{j}{\bf S}_{2j+1}\cdot {\bf S}_{2j+2}\,,
\label{eff}\eeq
with $J_{1}=4t_{in}^{2}/U$ and $J_{2}=4t_{out}^{2}/U$.  Since
experimentally $t_{in}$ and $t_{out}$ can be controlled independently
\cite{RGBDL}  by adjusting the intensities of the laser beams that
generate the superlattice, we will  assume that both $J_{1}$ and
$J_{2}$ can, in general, be {\it time-dependent} functions
$J_{1}(t), J_{2}(t)$. Additionally, we note that even though for
bosons  the sign of the coupling constants is normally  positive
(ferromagnetic interactions), experimentally it is also possible to
change the sign to be negative \cite{Bloch4}.

\subsection{Initial state}
The starting point  of our analysis is a system initially prepared
in an array of triplet pairs on the neighboring sites of a double-well superlattice,
\beq\label{instate}
&&|\psi(t=0)\rangle =\prod_{j}|t^z_{2j,2j+1}\rangle\,,\\
&&|t^z_{j,j+1}\rangle=\frac{1}{\sqrt{2}}(|\uparrow\rangle_{j}|\downarrow\rangle_{j+1}+|\downarrow\rangle_
{j}|\uparrow\rangle_{j+1})\,.
\eeq
This state has been recently realized  in the laboratory
\cite{Bloch4}. In this experiment,  after  first preparing a Mott
insulator  with two bosonic  atoms per double well,  the atoms  were
transferred into a triplet state configuration by using spin-changing collisions~\cite{widera-2005}.

\begin{figure}[ht]
\begin{center}
\includegraphics[width=0.45\textwidth]{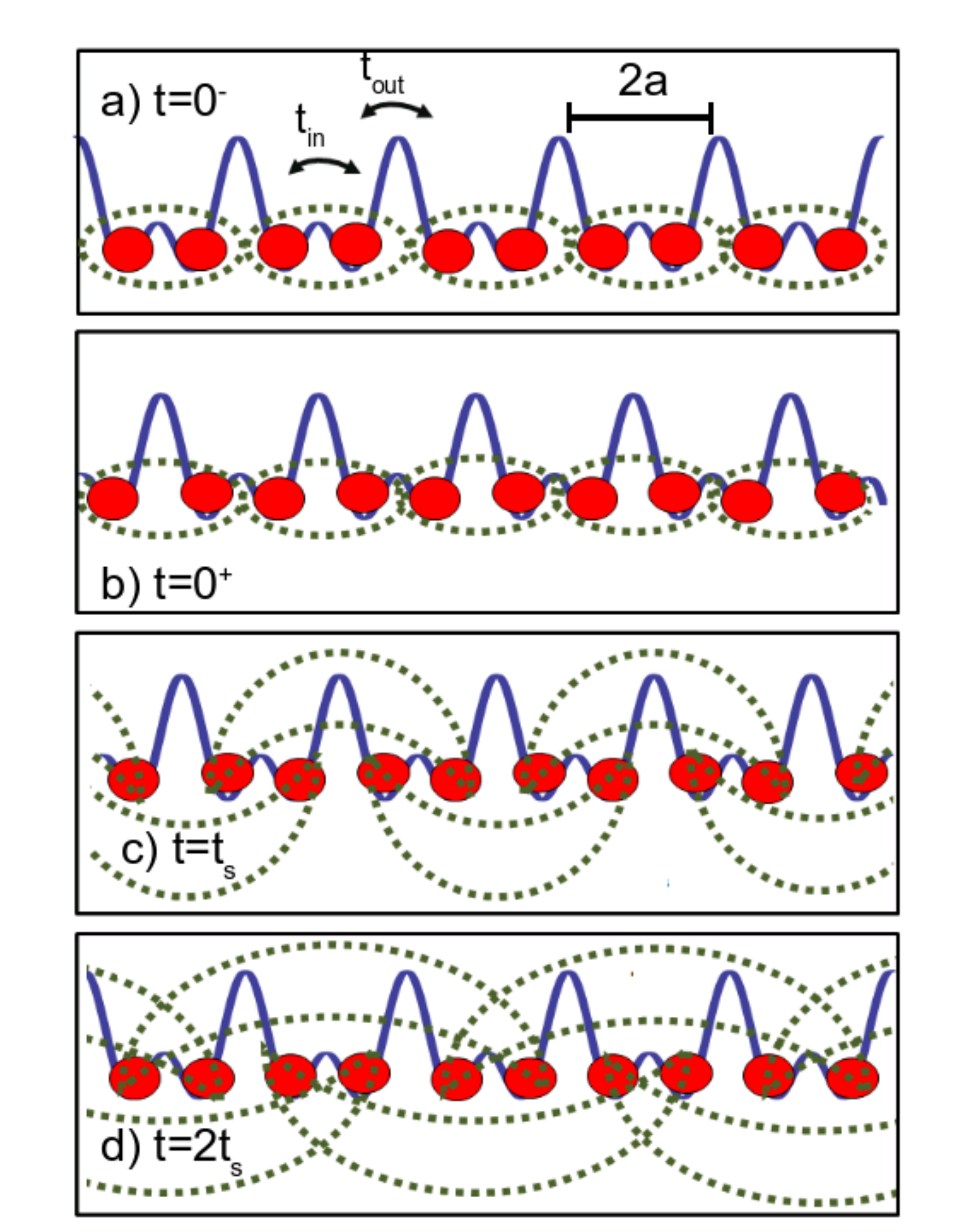}
\caption{\label{cartoon} \figpreamble a) The initial state in the
superlattice corresponds to a product of triplets at adjacent sites. $a$ is the lattice spacing.
b) At time $t=0^+$ the intra-well tunneling  is suppressed and the
inter-well tunneling is allowed. c) At $t=t_s$ the entanglement between
adjacent pairs is redistributed between pairs of length 3. d)
If the switching procedure is repeated, the entanglement
propagates to atoms separated by 5 wells and after $n$ switches
by $2n+1$ wells.}
\end{center}
\end{figure}
For the following, it is convenient to characterize the initial
state  as a {\it triplet valence bond} state of length 1. Although
this state is a ground state of the system of independent wells, it is
{\it not} an eigenstate of a system of coupled wells. Therefore,
changing the couplings $J_{1,2}$ at $t>0$ leads to a complicated
correlated dynamics. The specific time evolution depends
significantly on the ratio of the couplings $J_{1}$ and $J_{2}$.

\subsection{Switching procedures}
We consider and characterize in details three specific cases:
\begin{enumerate}
\item  Single switch: $J_{1}[t>0]=0, J_{2}[t>0]=J$.

\item  Periodic switch: $J_{1}[(2n+1) t_{s}>t>2 n t_{s}]=0,
J_{2}[(2n+1) t_{s}>t> 2 n t_s]=J$ while $J_{1}[ (2n+2) t_{s}>t> (2
n+1)t_{s}]=J, J_{2}[(2 n+2)t_{s}>t>(2n+1)t_{s}]=0$ with
$n=0,1,2,\dots$ and switching time $t_s$ specified below.

\item Homogeneous switch: $J_{1}[t>0]=J_{2}[t>0]=J$.
\end{enumerate}
The Hamiltonian in the first two cases consists of decoupled double
wells and allows a simple analytical treatment
(Sections~\ref{single_switch},\ref{many_switch}). The homogeneous
switch involves the complicated many-body dynamics of the Heisenberg
chain and will be analyzed using numerical tools
(Section~\ref{uniform}).

It is convenient to introduce the bond operators \cite{Sachdev}
which create singlet and triplet pairs at different bonds:
\beq
\hat{s}_{j,j+1}^\dagger|0\rangle &=&
|s_{j,j+1}\rangle=\frac{1}{\sqrt{2}}(|\uparrow\rangle_{j}
|\downarrow\rangle_{j+1}-|\downarrow\rangle_{j}|\uparrow\rangle_{j+1})\nonumber\,,\\
\hat{t}_{j,j+1}^{z\dagger}|0\rangle&=&|t^z_{j,j+1}\rangle \label{bond}\,,\\
\hat{t}_{j,j+1}^{x\dagger}|0\rangle &=&
|t^x_{j,j+1}\rangle=\frac{1}{\sqrt{2}}(|\uparrow\rangle_{j}
|\uparrow\rangle_{j+1}-|\downarrow\rangle_{j}|\downarrow\rangle_{j+1})\nonumber\,,\\
\hat{t}_{j,j+1}^{y
\dagger}|0\rangle&=&|t^y_{j,j+1}\rangle=\frac{i}{\sqrt{2}}(|\uparrow\rangle_{j}|\uparrow\rangle_{j+1}+|\downarrow\rangle_{j}|
\downarrow\rangle_{j+1})\nonumber\,
\eeq
($|0\rangle$ denotes the state with no atoms). These operators satisfy bosonic commutation relations
and the constraint
\beq
\sum_{\alpha=x,y,z}\hat{t}^{\alpha\dagger}_{j,j+1}
\hat{t}^{\alpha}_{j,j+1}+ \hat{s}^\dagger_{j,j+1} \hat{s}_{j,j+1}=1\,,
\eeq
which follows from the completeness of the Hilbert space of states
of an individual double well. We start our analysis by studying the
single switch dynamics.

\section{Single switch\label{single_switch}}

In the case $J_1=0$ and $J_2=J$, the evolution operator
$U(t)=e^{-itH/\hbar}$  can be written analytically. It is given by
\beq\label{uop} U_{odd}(t)=e^{iAt} \prod_{j}[\cos(\frac{J
t}{2\hbar}){\bf{1}}+ i\sin(\frac{J t}{2\hbar}) {\bf
{P}}_{2j+1,2j+2}]\,,\eeq
where ${\bf {P}}_{2j+1,2j+2}=(1/2)+2{\bf
S}_{2j+1}\cdot {\bf S}_{2j+2}$ is the swap operator (interchanges
the spins) at sites ${2j+1,2j+2}$ and $A$ is an irrelevant phase factor equal to $- J ( N+2)/(4 \hbar)$. From  Eq.~(\ref{uop}) it is clear that the
evolution is periodic with the period
\beq T&=& 2 t_s\notag\,,\\
J t_s&\equiv& \pi\hbar\,.
\eeq

At times $t=(2n+1)t_s$, the evolution operator reduces to a product
of the swap operators which, upon acting on the initial state,
distribute the entanglement from atoms at sites $(2j,2j+1)$ to atoms
at $(2j+1,2j+4)$, leading to the formation of a quantum state with
valence bond length equal to 3 (see Fig.1).
\beq
\label{vb3}
|\psi_{t=t_{s}}\rangle^{(1)}=\prod_{j}|t^z_{2j+1,2j+4}\rangle \,.
\eeq

\begin{figure}[ht]
\begin{center}
\includegraphics[width=0.45\textwidth]{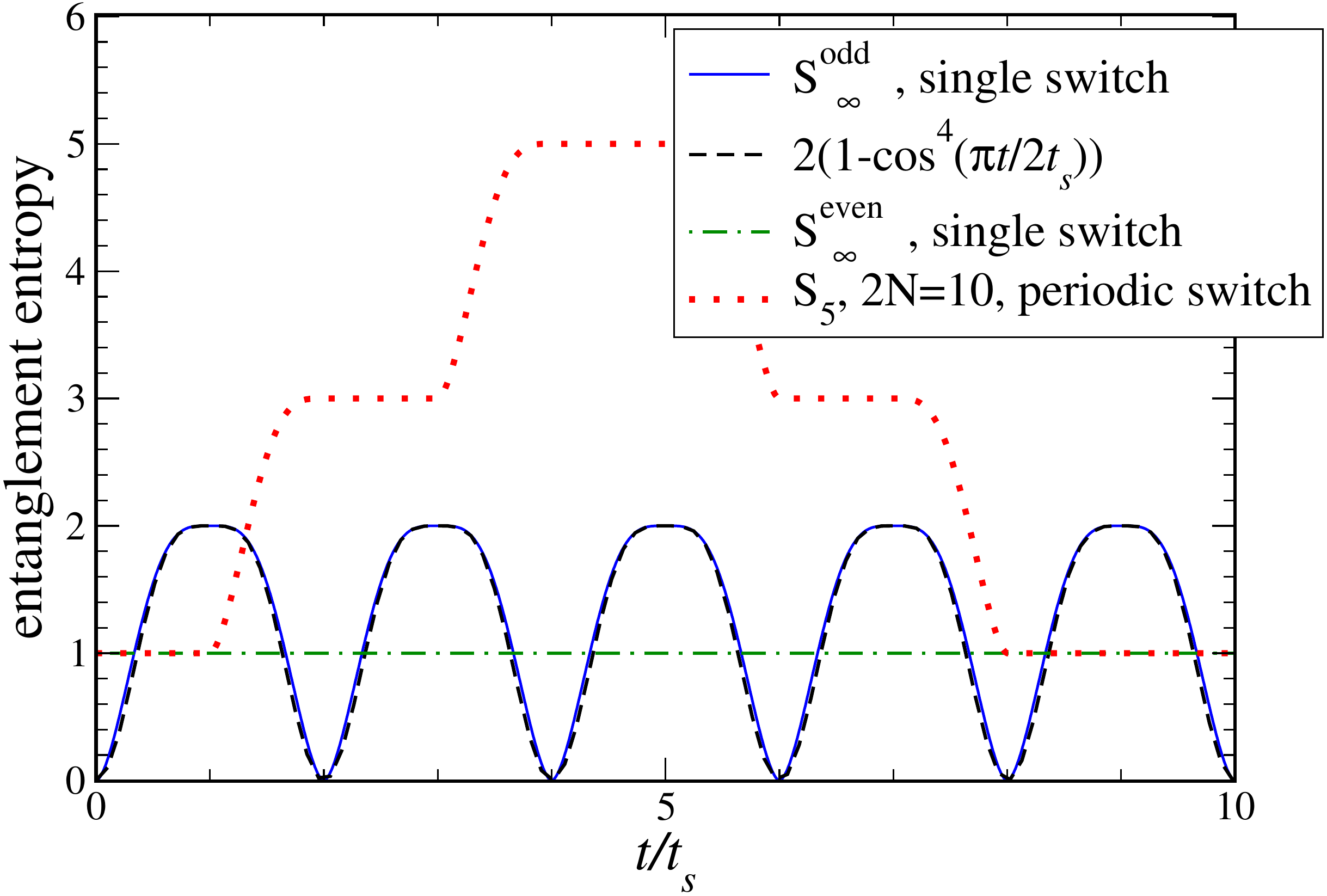}
\caption{\label{switchent} \figpreamble Entanglement entropy for the single
switch and for the periodic switch (numerical result, TEBD), $t_s=\pi\hbar/J$. We used an infinite lattice
for the former and one with $2N=10$ for the latter and calculated
the entanglement entropy for half of the chain. While for the
single switch the period is $2t_s$, for the periodic switch the
initial state is recovered after $t=2Nt_s$. The single switch is well described by expression~(\ref{cos4}).}
\end{center}
\end{figure}

The effect of this redistribution on the entanglement entropy
is shown on Fig.~\ref{switchent}. We observe that while for odd
bipartitions the entanglement entropy oscillates between 0 and 2,
for even bipartitions $S^{even}_{\infty}$ remains constantly 1.
This is consistent with the fact that for any state which can be
represented as a single valence bond state the entanglement entropy
is equal to the number of EPR-pairs shared by the subsystems
\cite{alet-2007} (in our case this number is 0 and 2 at
$nt_s$ for the odd bipartitions and 1 for the even). The
oscillation follows closely, but not exactly, the curve
\beq
\label{cos4}
S^{odd}_{\infty}(t)\approx 2(1-\cos^4(\frac{Jt}{2\hbar}))\,.
\eeq

The singlet and triplet populations at adjacent sites are quantities
that  can be experimentally  probed via singlet-triplet
spectroscopic measurements and Stern-Gerlach techniques~\cite{RGBDL}.
In terms of bond operators (see Eq.~(\ref{bond}))  these quantities are defined as:

\begin{eqnarray}
 t_{even}^{x,y,z}(t)&=&\frac{1}{N}\sum_j
\langle \psi(t)|{\hat{t}^{x,y,z\dagger}_{2j,2j+1}}
{\hat{t}^{x,y,x}_{2j,2j+1}}|\psi(t)\rangle\notag\,,\\
s_{even}(t)&=&\frac{1}{N}\sum_j \langle
\psi(t)|{\hat{s}^{\dagger}_{2j,2j+1}}
{\hat{s}_{2j,2j+1}}|\psi(t)\rangle\notag\,,\\
 t_{odd}^{x,y,z}(t)&=&\frac{1}{N}\sum_j
\langle \psi(t)|{\hat{t}^{x,y,z\dagger}_{2j+1,2j+2}}
{\hat{t}^{x,y,x}_{2j+1,2j+2}}|\psi(t)\rangle\notag\,,\\
s_{odd}(t)&=&\frac{1}{N}\sum_j \langle
\psi(t)|{\hat{s}^{\dagger}_{2j+1,2j+2}}
{\hat{s}_{2j+1,2j+2}}|\psi(t)\rangle\notag\,.\end{eqnarray}

Using the evolution operator~(\ref{uop}), the singlet-triplet
populations can be shown to evolve as
\beq
t^z_{even}(t)&=&\frac{1}{4}\left(1+3\cos^4\left(\frac{Jt}{2\hbar}\right)\right)\,,\\
t^{x,y}_{even}(t)&=&s_{even}(t)=\frac{1}{4}\left(1-\cos^4\left(\frac{Jt}{2\hbar}\right)\right)\nonumber\,,\\
t^{x,y,z}_{odd}(t)&=&s_{odd}(t)=\frac{1}{4}\,.\nonumber
\eeq

\begin{figure}[t]
\begin{center}
\includegraphics[width=0.45\textwidth]{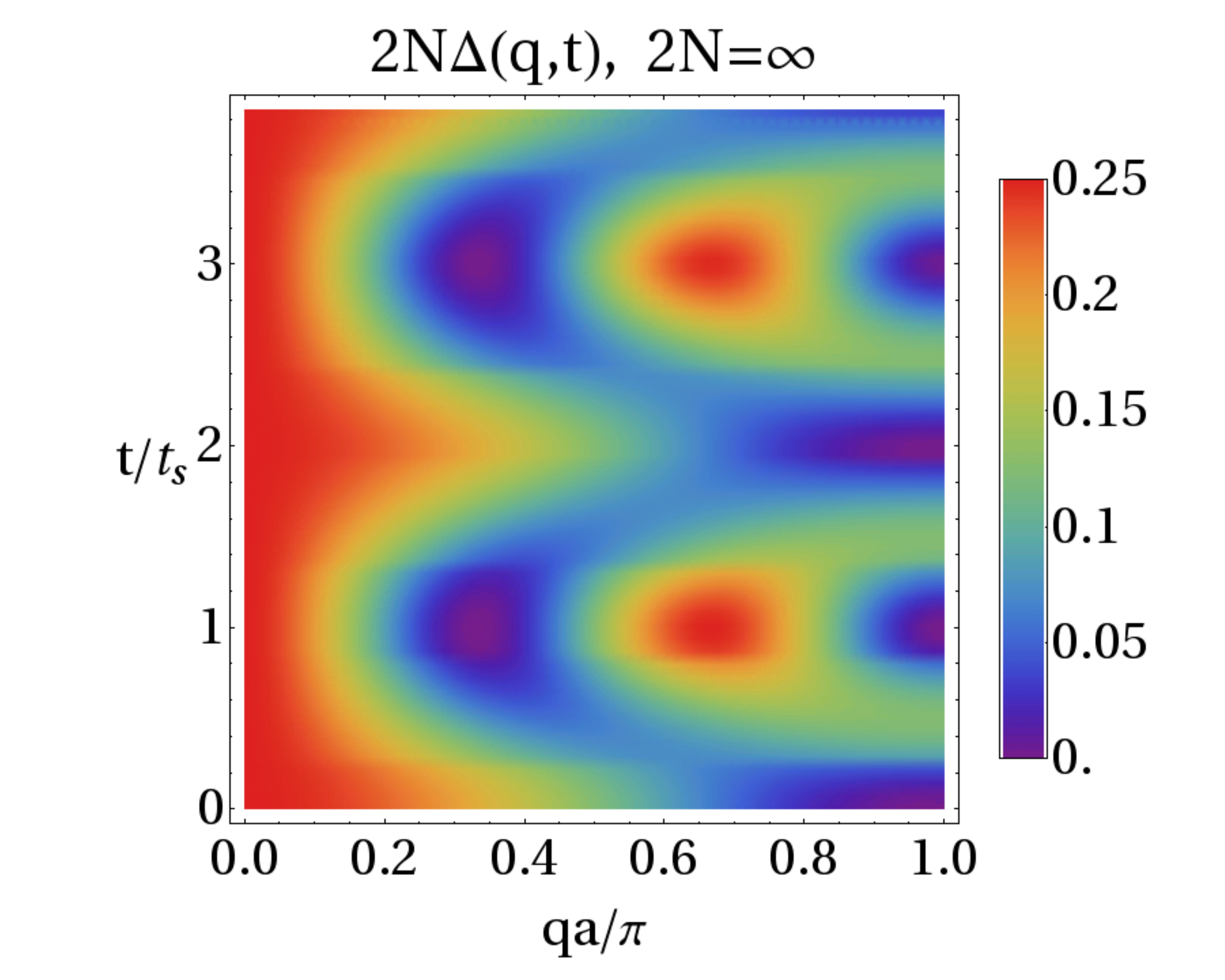}
\caption{\label{singleswitchnoise} \figpreamble The noise correlations for two periods in the
single switch, $t_s=\pi\hbar/J$.  Numerical TEBD simulation for the infinite lattice.}
\end{center}
\end{figure}

The coherence of the singlet-triplet oscillations can help to
characterize the quality of the dynamical evolution. These
measurements, however, are only local and do not give any indication
of the distance between the entangled atoms generated at $t =
(2n+1)t_s$. The latter, on the other hand,  can be probed by
measuring density-density correlations of the expanding cloud or
noise correlations~\cite{ADL}
\beq
G(Q(r),Q'(r'))=\frac{\sum_{\sigma\sigma'} \langle \hat
{n}_{Q(r)}^\sigma  \hat {n}_{Q'(r')}^{\sigma'}
\rangle}{\sum_{\sigma\sigma'} \langle \hat {n}_{Q(r)}^\sigma \rangle
\langle \hat {n}_{Q'(r')}^{\sigma'}\rangle} -1\,,
\eeq
where $\hat{n}_{Q(r)}^{\sigma}$ is the atom number operator for the
component $\sigma$ at position $r$ after time of flight.
$G(Q(r),Q'(r'))$ is directly related to the momentum-momentum
correlations of the atomic cloud at the release time, $t_{R}$.
Deep in the Mott insulator regime $G(Q(r),Q'(r'))$ can be
rewritten in terms of spin operators as
\beq\label{noise}
G(q,t_{R})\!&=&\!\langle\psi(t_R)|\frac{1}{2N^{2}}\sum_{i\neq
j}e^{iqa(i-j)}(\frac{1}{4}+{\bf S}_{i}\cdot{\bf S}_{j})|\psi(t_R)\rangle \notag \\
&=&\frac{1}{2}\delta_{q,0}+ \Delta(q,t_R)\,,
\eeq
where $q=Q-Q'$ and $a$ is the lattice spacing~(Fig. \ref{cartoon}). While the first term in Eq.~(\ref{noise})  reproduces
the interference peaks at reciprocal lattice vectors characteristic
of the Mott insulator state (due to the bunching of the bosons), the
second term $\Delta(q,t_R)$ provides additional information about
the spin order in the system. For example, if the system is released
exactly at times $t_R=n t_s$ when it is in a valence bond state of
length $l$ (here $l=1,3$), $\Delta(q,t_R)$ will exhibit spatial
oscillations with periodicity dictated by the distance between the
entangled atoms (see Fig. \ref{singleswitchnoise})
\beq \label{sfactor}
\Delta(q,t_R=n t_s)=\frac{1}{4N}[1+\cos(qal)]\,.
\eeq

We note that the factor $N$ in the denominator originates from the
short-range character of the interactions and therefore the
entanglement is only shared between pairs. It limits the
applicability  of noise correlations as a suitable experimental
probe  in systems with large number of atoms. However, the $1/N$
factor should not be a problem in  current $1D$   systems  with
approximately  20 atoms per tube \cite{paredes}.

\section{periodic switch\label{many_switch}}

\subsection{Generic case}
\begin{figure}[b]
\begin{center}
\includegraphics[width=0.45\textwidth]{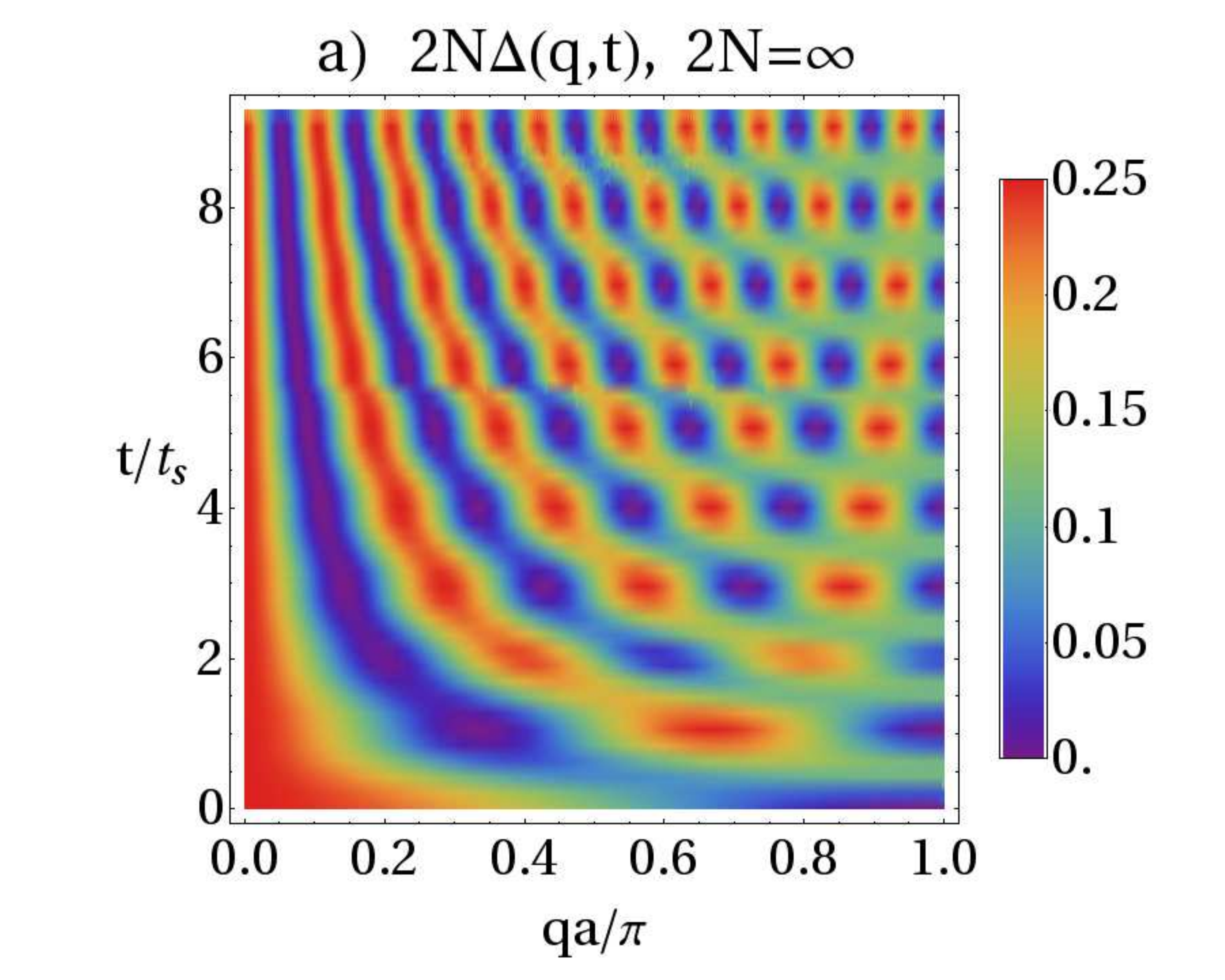}
\includegraphics[width=0.45\textwidth]{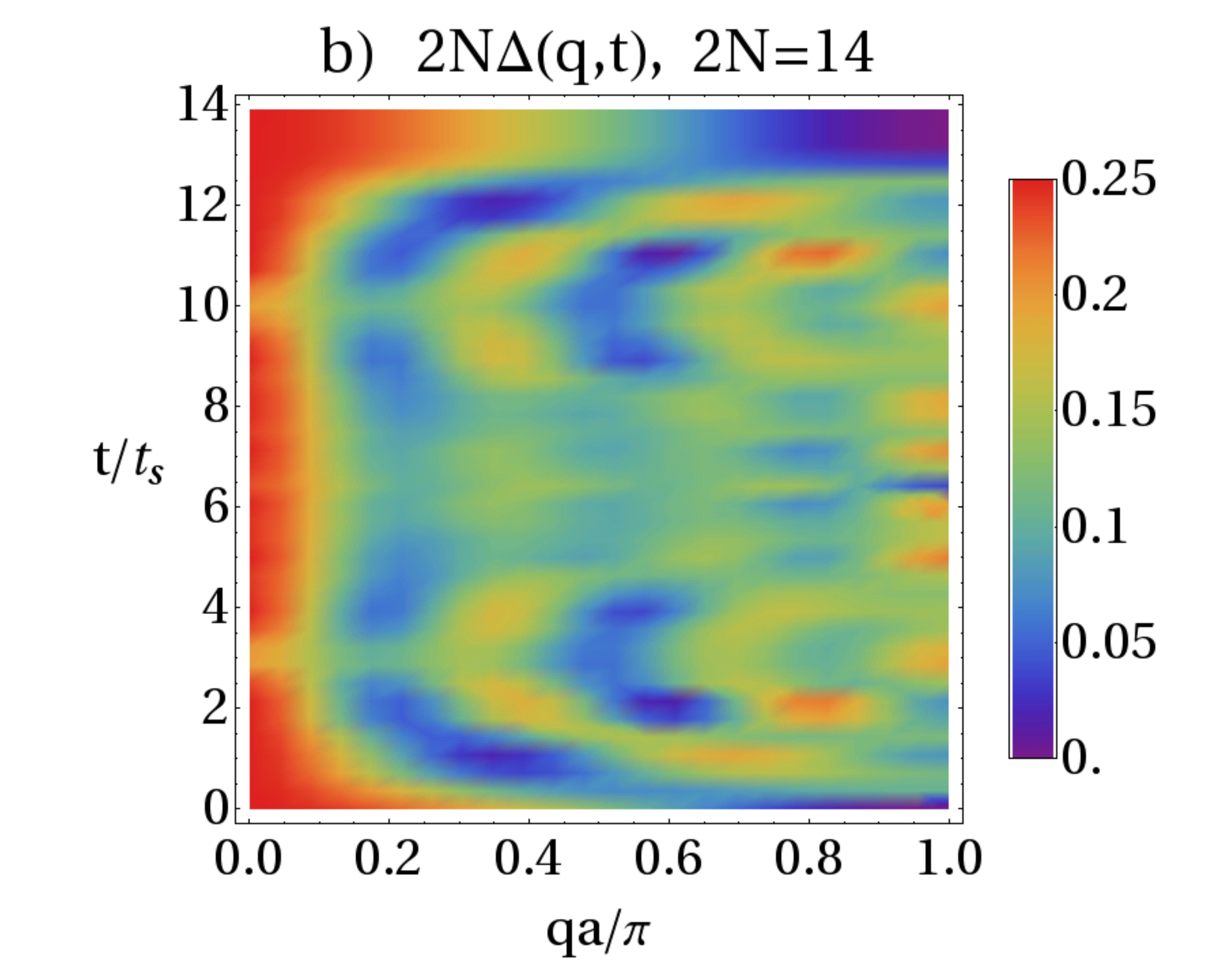}
\caption{\label{switchnoise} \figpreamble The noise correlations during the
periodic switch, $t_s=\pi\hbar/J$. a) TEBD Simulation in the limit $N\rightarrow \infty$. The fact that all the entangled pairs are of the same length is reflected in the periodic pattern. b) Exact Diagonalization on an open chain, $2N=14$. The superposition of the triplet valence bonds with different lengths
in the intermediate state around $t/t_s=N-1$ leads to a very weakly
structured signal.}
\end{center}
\end{figure}

We now consider the iterative sequence of switching off and on  the
couplings $J_{1}, J_{2}$ every $t_{s}$. One might think that if at
time $t_s$ one reverses the couplings from $J_{1}=0, J_{2}=J$ to
$J_{1}=J, J_{2}=0$, the dynamics will just return the state  into
its original form, i.e. from  Eq.~(\ref{vb3}) to Eq.~(\ref{instate});
however, this does {\it not} happen. On the contrary, as a result of the evolution under the
swap operators, atoms separated by {\it four} lattice sites become now connected by triplet valence bonds and so,
at the time $t=2t_{s}$, the state evolves into a valence bond state with $l=5$
(Fig.~\ref{cartoon}),
\beq
|\psi_{t=2t_{s}}\rangle^{(2)}=\prod_{j}|t^z_{2j,2j+5}\rangle\,.
\eeq
The successive repetition of the switching procedure leads to the
propagation of the entanglement across the lattice and after
$n$ switches,  performed at times $ k t_{s}$ ($k=1,\dots n$), one
obtains entangled pairs with length $2n+1$.

In the experimentally relevant case of an open chain, the sequential
incrementation of the length of the entangled pairs is stopped when one member
of the pair reaches the boundary of the lattice. The pair is then
reflected and continues moving through the lattice with its length remaining unchanged.
Consequently, when  after $N-1$ switches the pair initially
located at the center of the lattice reaches the boundary, a
particular state  that has the maximal possible length of
entangled pairs is formed. While for  an odd number of double wells
it corresponds to a  state with an EPR-pair connecting the edges of
the chain,
\beq
\notag|\psi_{t=t_s(N-1)}\rangle=\tz{0}{2N-1}\!\!\!\!\!\prod_{\scriptscriptstyle j=1}^{\scriptscriptstyle(N-1)/2}\!\!\!\!\tz{2j-1}{2N-2j-1}\tz{2j}{2N-2j}\,\!,
\eeq
 for even $N$ the maximal length of entangled pairs is
$l=2N-1$,
\beq
|\psi_{t=t_s(N-1)}\rangle=\prod_{\scriptscriptstyle j=0}^{\scriptscriptstyle(N-2)/2}\tz{2j}{2N-2j-2}\tz{2j+1}{2N-2j-1}\,.\notag
\eeq
Since the entanglement entropy of the state partitioned into its
left and right half is simply given by the number of EPR-pairs
connecting the two parts~\cite{alet-2007}, the state $|\psi_{t=t_s(N-1)}\rangle$ has
maximal entanglement entropy $S_N=N$.  This growth of the
entanglement for the case $2N=10$ is depicted in Fig.~\ref{switchent}.

\subsection{Implementation of remote entanglement protocol}
As we have seen, by applying the iterative swapping procedure to an
open chain it is possible to engineer a state which has  maximally
separated entangled atoms and largest bipartite entanglement. Such a state  can have relevant application in lattice-based quantum
information proposals due to its large persistency of entanglement
because in this case $N$ qubits have to be measured to disentangle
the state. The persistency of entanglement quantifies the robustness
of the entanglement to noise. We remark that in this respect a cat state (macroscopic quantum
superposition state e.g. $\frac{1}{\sqrt{2}} (|\uparrow \uparrow \dots
\uparrow \uparrow \rangle + |\downarrow \downarrow \dots  \downarrow
\downarrow \rangle)$ ) is fragile as even a single local measurement is sufficient to reduce
it to a product state. The state we are engineering has 
persistency of entanglement as large as that of a cluster state, which is
one of the key prerequisites for using it as a one-way quantum
computer resource \cite{Briegel}.

Moreover, the $|\psi_{t=t_s(N-1)}\rangle$ state is an eigenstate of
the $N$-th switching operator,
$|\psi_{t=(N-1)t_s}\rangle=|\psi_{t=Nt_s}\rangle$ and thus after
$2N$ switches the state will be   rolled back to  the initial
nearest-neighbor triplet-product state. This  property can be useful
for experimentally probing the state and quantifying the fidelity of
the procedure. For example, by measuring the quality of the triplet
product state after $2N$ switches one can get information about
errors that occurred during the swapping process.

We also note that even though we focused on the case of an
initial array of triplet pairs, similar considerations  hold if
instead  of triplets one starts with singlets or changes the sign
of the coupling constants (as it would be in the case of fermions).

In addition, our swapping procedure can also be used for
transporting a particular state of an atom without directly moving the
particles: If we initially prepare all the  atoms in the same state,
say $\downarrow$, except for the atom at site $i$ which we prepare
in state $\uparrow$, after $n$ periodic switches the state $\uparrow$
will be transferred to the atom located at site $i+n$.

The  long-range entanglement produced by the switching procedure can
be experimentally probed  by noise-correlation measurements.
Although for finite lattices the expected ideal pattern of well-defined peaks   at $t=nt_s$ (see Eq.~(\ref{noise})) changes to one with  less regular structure due to the distribution of
different valence bond lengths, Fig.~\ref{switchnoise} shows that it
still contains relevant information such as the formation of well-defined peaks at $q=0$ and $q=\pi/a$ when the distance between
entangled atoms becomes maximal.

\subsection{Non-ideal conditions}

Up to this point  we have assumed that Eq.~(\ref{eff}) accurately
describes the many-body dynamics. However, defects such as holes or
doubly occupied sites will make this assumption invalid.

We should emphasize that there is one particular condition which
makes the entanglement generation possible despite the presence of holes. Namely, this occurs when the single-particle tunneling
time is engineered to be commensurate with $t_s$. However, if this condition is not satisfied, in general the holes will hinder the generation of
long-distance entangled pairs and they should be suppressed for
example by implementing additional filtering schemes such as the one
proposed in Ref.~\cite{Rabl}.

Additionally, even though Eq.~(\ref{eff}) was
derived by taking  into account only virtual particle-hole
excitations, real particle-hole excitations will certainly take place
during the dynamical evolution. They would lead to oscillations on
top of the effective Hamiltonian dynamics with amplitude $J/U$ and
periodicity $\sim h/U$. Therefore, in order to efficiently average them
out one has to work in the strongly correlated regime, i.e. with the condition $t_{in,out} \ll U$, though this implies smaller time scales for the
dynamical evolution. In typical experiments, working in a  parameter
regime where particle-hole excitation effects are negligible requires a superexchange coupling $J/h$
of the order of 1 kHz ($t_{s}\sim $ 1 ms) and thus for a system with
approximately $20$ lattice sites, it will take about $10$ ms to
generate entanglement between the atoms at the edges of the
cloud.

Another aspect of our procedure is that the long-distance entangled
pairs are generated by switching the interactions at specific
moments of time. In practice however one always expects switching time
uncertainties $\delta t$ and therefore the interval between consecutive
steps will not be exactly $t_s$ but $t_{s} +\delta t$. Such inaccuracies
will accumulate and will degrade the quality of the final state
exponentially with the number of lattice sites and the number of
switches made during the process. Defining the fidelity of a state as
$\mathcal{F}=|\langle \psi _{t=nt_s} | \psi
_{t=nt_s}^{ideal}\rangle|^2$, where $|\psi _{t=nt_s}^{ideal}\rangle$
and $|\psi _{t=nt_s}\rangle$  are the ideal and actual states
generated after $n$ iterations, one can estimate the degradation of
fidelity using Eq.~(\ref{uop})
\beq
\mathcal{F}\sim \mathcal{F}_0(1-\frac{n (\delta t)^2 N}{4})
\eeq
where $\mathcal{F}_0=|\langle \psi(t=0) |\prod _j
|t^z_{2j,2j+1}\rangle|^2$ is the fidelity of the initial state.

\subsection{Entanglement purification }

To overcome all the limitations mentioned above one can combine our
periodic switching scheme with entanglement purification protocols.
Starting from a large ensemble of generated low-fidelity entangled
pairs, these protocols distill a smaller sub-ensemble which has
sufficiently high fidelity. Entanglement purification can be
implemented  in a spin-dependent 2D superlattice as follows: after
creating an array of $1D$ independent chains along
$x$-direction by suppressing tunneling along $y$-direction, one
can use our procedure to generate many parallel long-distance
entangled pairs within the 1D chains, i.e. an atom at the site $(i,j)$
will be entangled with one at $(i+l,j)$. Then tunneling along the
$x$-direction should be inhibited and the following iterative
procedures  be applied:

\begin{enumerate}
  \item Lower the intra-well barriers along the $y$-direction of a
spin-dependent superlattice, allowing  only one of the species  to
tunnel~\cite{andre-2002}. This will introduce Ising-type interactions
$\sum_j J'\bf{S}_{i,2j}^z \bf{S}_{i,2j+1}^z$ between atoms at
adjacent sites along $y$-axis and therefore will couple entangled pairs
at $(i,2j)$ - $(i+l,2j)$ with pairs at $(i,2j+1)$ - $(i+l,2j+1)$
respectively. 
  \item Combine the Ising interaction with single-particle
rotations, realized with the help of external magnetic fields, to
implement the C-Not gate required for the purification schemes
described in Ref. \cite{Bennett}.
  \item Measure the spins at the $(i,2j)$ and $(i+l,2j)$  wells. If they turn out to be parallel,  keep the corresponding pair at $(i,2j+1)$ and $(i+l,2j+1)$, otherwise discard it.
  \item Release the  measured atoms and merge
the $(i,2j)$ and $(i,2j+1)$ wells into a single one. Repetition of
the above protocol will distill from the low-fidelity pairs the ones 
with higher fidelity.
\end{enumerate}

Let us now briefly discuss the experimental realizability of such
purification protocols. To date, one of the main problems is the
experimental implementation of step 3 due to the difficulty of
measuring individual states at adjacent lattice sites. These atoms
are separated by a  distance of the order of an optical wavelength
and therefore diffraction fundamentally limits individual
addressability. One advantage of our scheme is that the
atoms in a pair that should be measured are in general separated by many
lattice sites, but nevertheless when the measurement is performed on one of the pairs,  nearest neighbor atoms are still affected. One possibility to overcome
this problem has been proposed recently in Ref.~\cite{gorshkov-2007}
where the use of  nonlinear atomic response  has
been suggested for coherent optical far-field manipulation of quantum systems with resolution of up to a few nanometers. The
implementation of the proposals of this kind in the controlled lattice
environment may allow proof-of-principle experimental
demonstration of quantum purification ideas.

\section{Homogeneous switch\label{uniform}}
An interesting question which arises from the dynamics of the
periodic switch is what happens with the quantum state if the double
wells are no longer decoupled completely, but instead there exists
a finite tunneling between them. One expects that in this case the
propagation of valence bond states will be suppressed after some period of
evolution. To address this question, in this section we consider the
case of a homogeneous switch $J_1/J_2=1$ (case (3) in our
classification), which formally can be considered as a particular
case of quench dynamics:  we prepare the system in a ground state of
one Hamiltonian -- a triplet (singlet)  product state -- and then
suddenly change the quantum Hamiltonian to a new one -- the isotropic
ferromagnetic Heisenberg Hamiltonian --, which  determines the
subsequent evolution.

In contrast to the periodic switch evolution, whose general
characteristics are independent of the singlet or triplet nature of
the starting  state, the dynamics of the homogeneous switch is strongly
affected by the symmetries of the initial state. Consequently, we consider the cases  when the initial state is in a triplet (singlet) configuration  separately. However, before starting the
discussion we first  provide a general overview of the
dynamics of quantum quenched systems.

\subsection{Quantum quench: general discussion\label{general}}

The time evolution of a quantum state after a quantum quench has
recently attracted a lot of theoretical
\cite{CC-all,CC:Entanglement,cramer-2007} and experimental
\cite{Bloch3,paredes,bloch-2007,Weiss} interest, in part due to the
possibility of varying  in real time the parameters of the optical lattice. For example, low-dimensional systems prepared in a
gapless state and subsequently quenched into an insulator state have been
experimentally studied,  addressing   questions such as relaxation
to thermal states and collapse and revival effects. The dynamics
of exactly solvable models, e.g. an  Ising chain
\cite{barouch-all,sengupta-2004,cazalilla}, have also been the topic of
investigation due to the fact that these systems satisfy many
conservation laws which lead to non-trivial equilibration
phenomena. Such behavior has been attempted to be explained in terms of
a generalized Gibbs ensemble \cite{Rigol:HCBosons}. From the
numerical side, recent advances in time-dependent density matrix renormalization group (DMRG) and TEBD methods ~\cite{daley:tdmrg,Vidal:TEBD,white-feiguin} have allowed to study the quantum dynamics in bosonic and fermionic 1D
systems~\cite{manmana-2007,kollath-2007,laeuchli-2008}. The
numerical simulations seem to support the absence of thermalization,
however, these methods  are  restricted to small and intermediate time scales. The
case of the quench from the gapped phase into the critical regime
has been studied using conformal field theory by P. Calabrese and J.
Cardy~\cite{CC-all,CC:Entanglement}. Numerical calculations
\cite{ChiaraMontangero,laeuchli-2008} support their results. The quench dynamics
between gapped states can also be attacked using methods of exact
solutions \cite{GDLP} and also demonstrate interesting dynamics
associated with the absence of thermalization. On the other hand, more
conventional approaches based on perturbative methods
\cite{gasenzer-2006} and diagrammatic expansions
\cite{rey-2004,gasenzer-2007} inevitably show dynamics associated
with thermalization scenario.

In the present work we adopt a numerical approach to deal with the
quantum quench  dynamics  and postpone  the analytical treatment
for future publications.

\subsection{Initially prepared triplet state}
Let us first consider the case  of the homogeneous switch dynamics
when the initial state is a product of triplet states
(Eq.~(\ref{instate})). In order to gain a general understanding of this
system, we note that while the initial state has broken rotational
and translational  symmetries, the Hamiltonian at $t>0$
(ferromagnetic Heisenberg) possesses both of these symmetries.
Although its low-energy excitations are dominated by the spin-wave
Goldstone modes corresponding to the broken {\it continuous}
(rotational) symmetry, the quantum dynamics involves a bunch of highly
excited modes which know nothing about the spontaneous breakdown of
the continuous symmetry. We therefore face a {\it dynamical} competition
between the initial state with broken symmetries associated with the
initial condition on the one hand, and the whole spectrum
reflecting both of these symmetries on the other. As a
result of this competition we expect the emergence of a complex
magnetic state and the growth of the entanglement entropy.

As we have pointed out, for the  correct description of quantum
dynamics it is not sufficient to rely on a low-energy effective theory
because the details of the spectrum can play a significant role. On the
other hand, if we start with a state which involves many excited
states, the characteristic features of the dispersion relation of the
low-energy modes can be not so important. Also, quantities studied
below are invariant under time-inversion symmetry and therefore the
dynamics of our problem should have the same common features as that of the antiferromagnetic Heisenberg model. As a result, some common
mechanisms should define the generic features of the quantum dynamics of
these models. It has been pointed out recently
\cite{CC-all,CC:Entanglement} that this generic behavior can
be understood in terms of classically moving quasiparticles
\cite{CC:Entanglement,cramer-2007}, whose transport correlations are
bounded by the light cone ({\it horizon effect}). We interpret our
results on the basis of these ideas.

\subsubsection{Entanglement}
\begin{figure}[t]
\begin{center}
\includegraphics[viewport=30 0 900 500,scale=0.35]{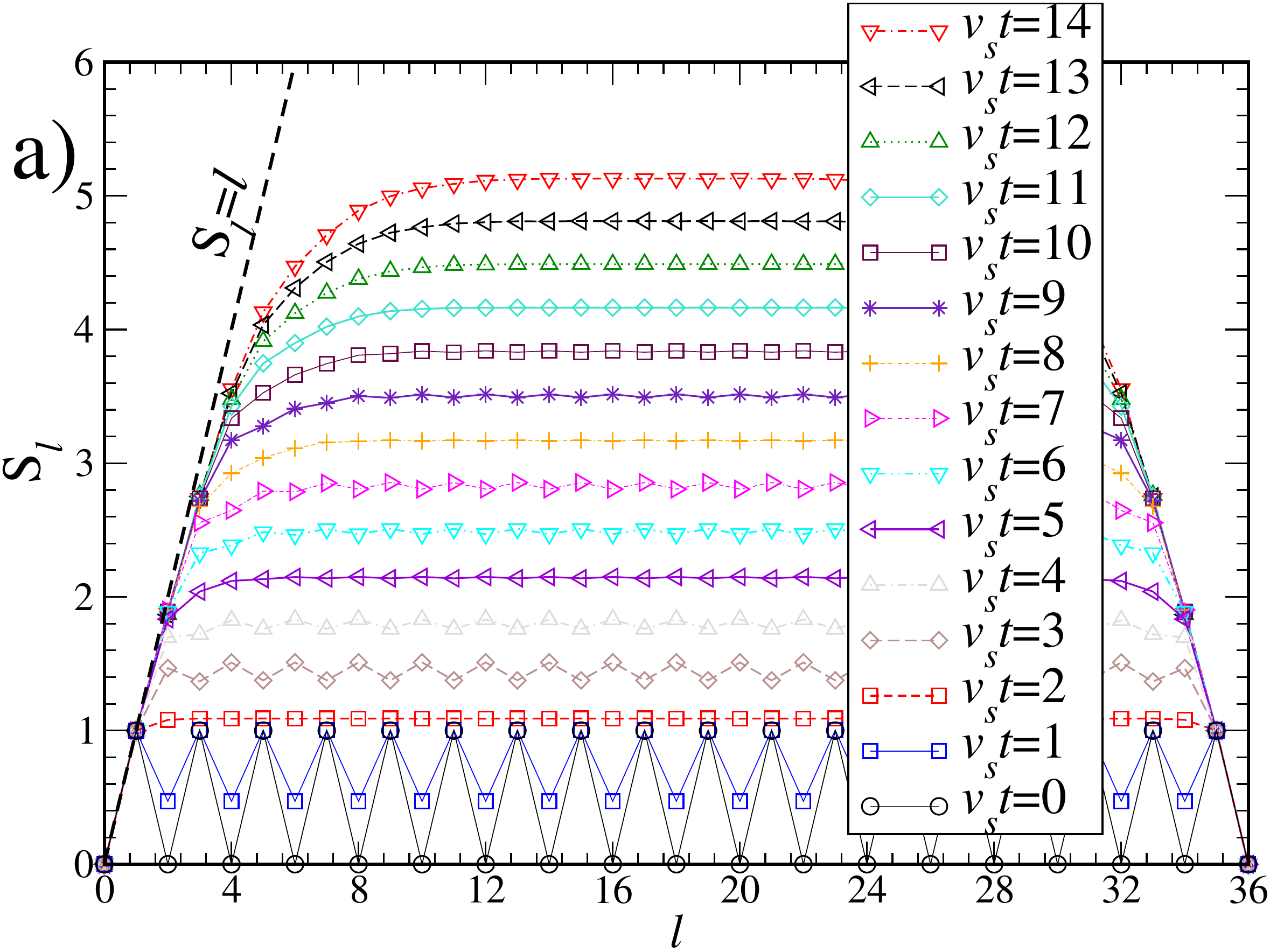}
\includegraphics[viewport=30 30 900 500,scale=0.35]{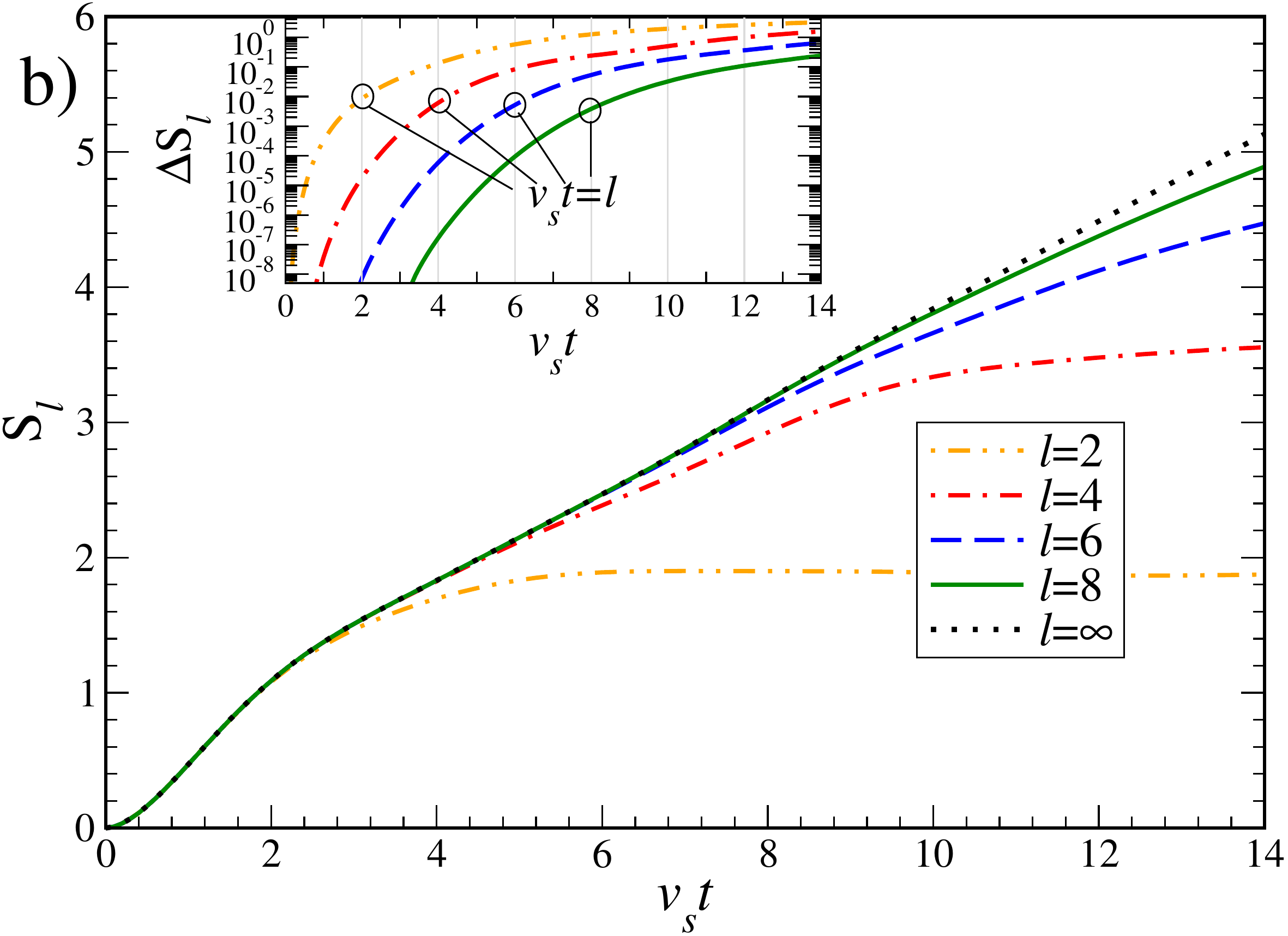}
\caption{\label{ent} \figpreamble The entanglement entropy for the homogeneous
switch, $v_s=J\pi/2\hbar$.  a) TEBD simulation for $2N=36$ with open boundary
conditions. $S_l$ approaches the line $l$. b) The crossover from
linear growth to saturation. Inset: deviation of the finite-block
entanglement entropy from the infinite value. The crossover is well characterized by the saturation time
defined by the spin-wave velocity, $t^*=\frac{l}{v_s}$.}
\end{center}
\end{figure}

We first focus on  the evolution of the entanglement. The spatially
anisotropic and weakly entangled initial state evolves into a highly
entangled state with restored translational symmetry. This behavior
is signaled by the growth of the entanglement entropy and the rapid
decay of the oscillations  between  even and odd bipartitions. 
In Fig.~\ref{ent} we plot the entanglement entropy of  blocks of
different sizes in a finite lattice. The plot shows that for short
times, after the recovery of translational invariance, the finite-block entanglement entropy exhibits linear growth. A saturation
to a value close to the maximal $S_l=l$ occurs for longer
times. This is in agreement with results obtained with the use of conformal field theory
\cite{CC:Entanglement} that predict a saturation value proportional
to $l$. The growth of the entanglement limits the applicability of
the numerical method (TEBD), as reasonable matrix dimensions (e.g.
$\chi=1000$) are only valid for weakly entangled systems ($S_l\leq
\log_2(\chi)\sim 10$). Consequently,  it is impossible to verify the
exact behavior of the entropy for large blocks. However, since in
the intermediate time regime the  dynamically evolved state in
finite lattices does not show significantly lower entanglement
compared to an infinite system, to study this regime one can make
simulations directly in the infinitely extended periodic system,
where the translational symmetry can be exploited. This allows to
reduce the computational cost by a factor of $N$ compared to the finite-lattice simulations \cite{Vidal:ITEBD}.

We study the crossover that takes place from the 'linear'-growth regime
where $S_l=S_\infty$ ($S_\infty$ stands for $S_\infty^{even}$, $S_\infty^{odd}$, for even and odd $l$ respectively), to a saturation towards a constant value. It is probed by the quantity (see Fig.~\ref{ent})
\beq
\Delta S_l &=& S_{\infty} -S_l.
\eeq
This crossover is a direct manifestation of the horizon effect. In
the case of conformal invariance, where relativistic dispersion
relation $\omega_k=v|k|$ is assumed, the distance between entangled
atoms is always smaller than $2vt$.  The entanglement grows linearly
as long as the horizon is smaller than the size of the block.
For the open chain considered here, with the block situated at
one of the edges, the horizon has to be twice as large as the block
length \cite{ChiaraMontangero}. This allows to define a crossover
time $t^*=l/v$ when $S_l[t>t^*]$ becomes a constant
\cite{CC:Entanglement}.  Fig.~\ref{ent} shows that using the
spin-wave velocity of the Heisenberg ferromagnet, $v_s=J\pi/2\hbar$, the
crossover indeed takes place around $t^*=l/v_s$. However, comparing
results of Fig.~\ref{ent} with the results of the quantum quench in
the $XXZ$-chain \cite{ChiaraMontangero}, we find that the crossover
in our case is much slower than in this system. The reason for this
is that in a one-dimensional lattice model the sharp crossover is
smeared out by lattice effects (which explain why $S_l<S_\infty$ even for $t<l/v_s$) and, more importantly, by  the  non-linear dispersion relation. Due to the latter, particles moving slower than $v_s$ have to
be taken into account, what results in a slower saturation of
$S_{l}(t)$ to a constant value at $t>t^*$.

While long-range effects at the 'horizon' are determined by the 'fast' spin waves and the results from conformal field theory are applicable, the slow quasi-particles will be of great importance for understanding the effects related to short-range phenomena.

\subsubsection{Singlet-triplet population}

\begin{figure}[t]
\begin{center}
\includegraphics[scale=0.35,angle=0]{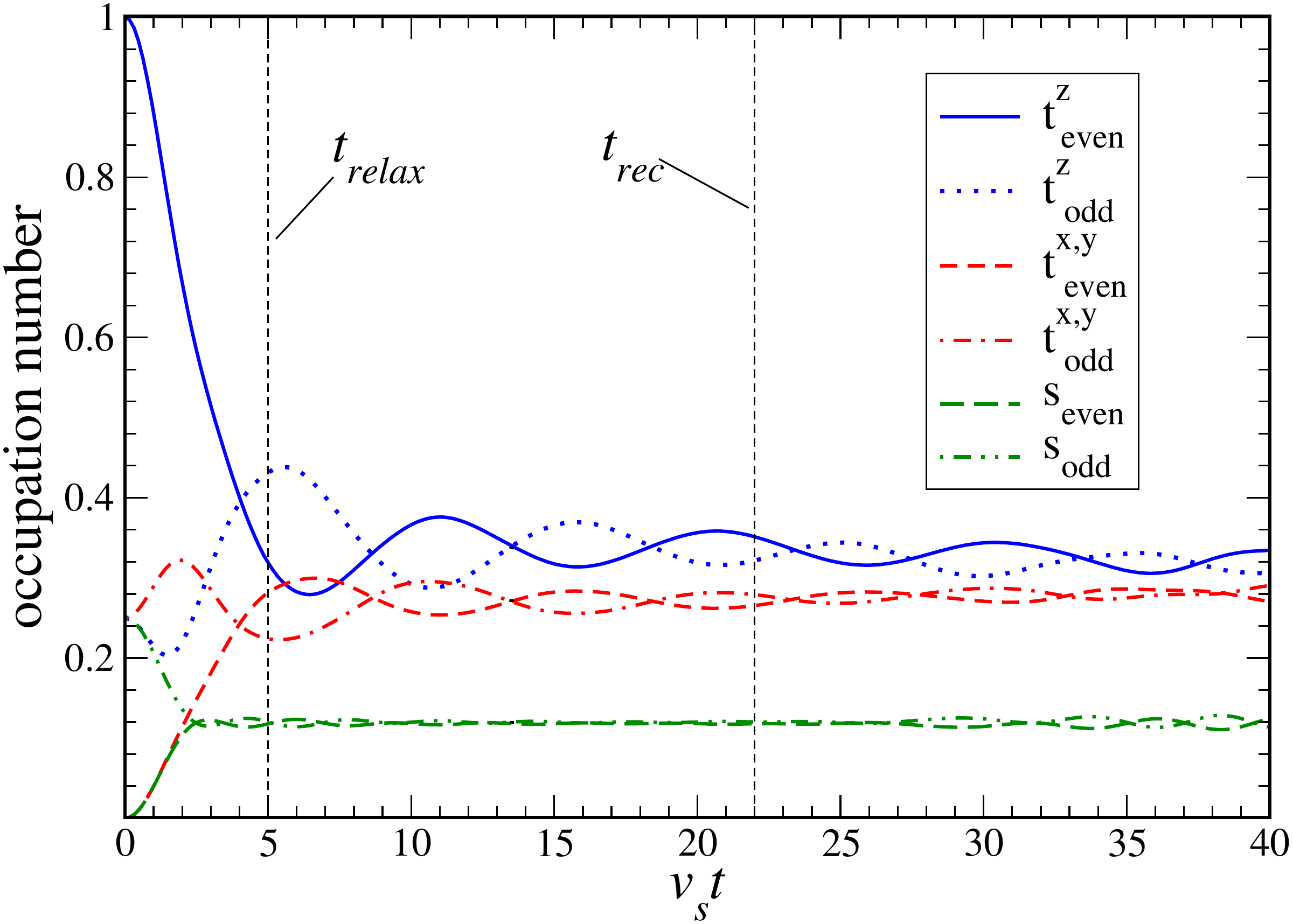}
\caption{\label{occ} \figpreamble The experimentally measurable triplet ($t^{x,y,z}$, see
text) and singlet ($s$) occupations at adjacent lattice sites, $v_s=J \pi/2\hbar$. Exact diagonalization for 2N=22 sites. The
equilibration of $t^z$ and $t^{x,y}$ is not complete. The oscillations of even and odd bond
correlations around the same value signal the recovery of the
translational symmetry.}
\end{center}
\end{figure}

To study further the dynamical relaxation and the recovery of broken
symmetries, we plot in Fig.~\ref{occ} the singlet-triplet population
at adjacent sites $(j,j+1)$. The data are obtained by using an unbiased exact
diagonalization technique (Lanczos algorithm \cite{Hochbruck:Lanczos}) on an open chain with
$2N=22$ sites. After a certain time interval $t_{relax}$ one expects that the
quenched initial values decay into a quasistationary regime, which
is destroyed at $t>t_{rec}=4N/2v_s$ due to quantum recurrence, when the
edges of the expanding light cone begin to interfere~\cite{cramer-2007}. 
From Fig.~\ref{occ} we determine that $v_s t_{relax} \approx 5$; the value $v_s t_{rec}=22$ corresponds to a conservative lower bound of the recurrence time. In the quasistationary regime only oscillations around an average value
are observed. These oscillations can be associated with the finite bandwidth of quasiparticle energies
\cite{CC-all}. We define average values for an arbitrary operator
$O$ in the quasistationary state as follows:
\beq
\langle O \rangle_{qs}=\frac{1}{t_{rec}-t_{relax}}\int_{t_{relax}}^{t_{rec}}dt \langle O(t)\rangle\, .
\eeq
Comparing the
$x,y,z$-triplet populations, we see a tendency of relaxation towards proximate values:
\begin{eqnarray*}
&\langle t^z_{even}\rangle_{qs}&=0.334\,,\\
&\langle t^z_{odd}\rangle_{qs}&=0.343\,,\\
&\langle t^{x,y}_{even}\rangle_{qs}&=0.267\,,\\
&\langle t^{x,y}_{odd}\rangle_{qs}&=0.272\,.
\end{eqnarray*}
The relaxation  of the singlet-triplet occupation numbers at even
and odd bonds towards the same quasistationary values indicates the
restoration of the translational symmetry, also suggested by the entropy
calculations. The difference between $t^z$ and $t^{x,y}$
of about $0.06$ implies that the rotational symmetry is not
completely restored. This difference is stable for various lattice
sizes and choices of $t_{relax}$ (which by definition allows a certain freedom in its choice). This is a direct indication of missing {\it
thermalization} in the quasistationary regime. The dynamic state does not fully reflect
the symmetries of the Hamiltonian.
\subsubsection{Structure factor}

\begin{figure}[ht]
\begin{center}
\includegraphics[width=0.45\textwidth]{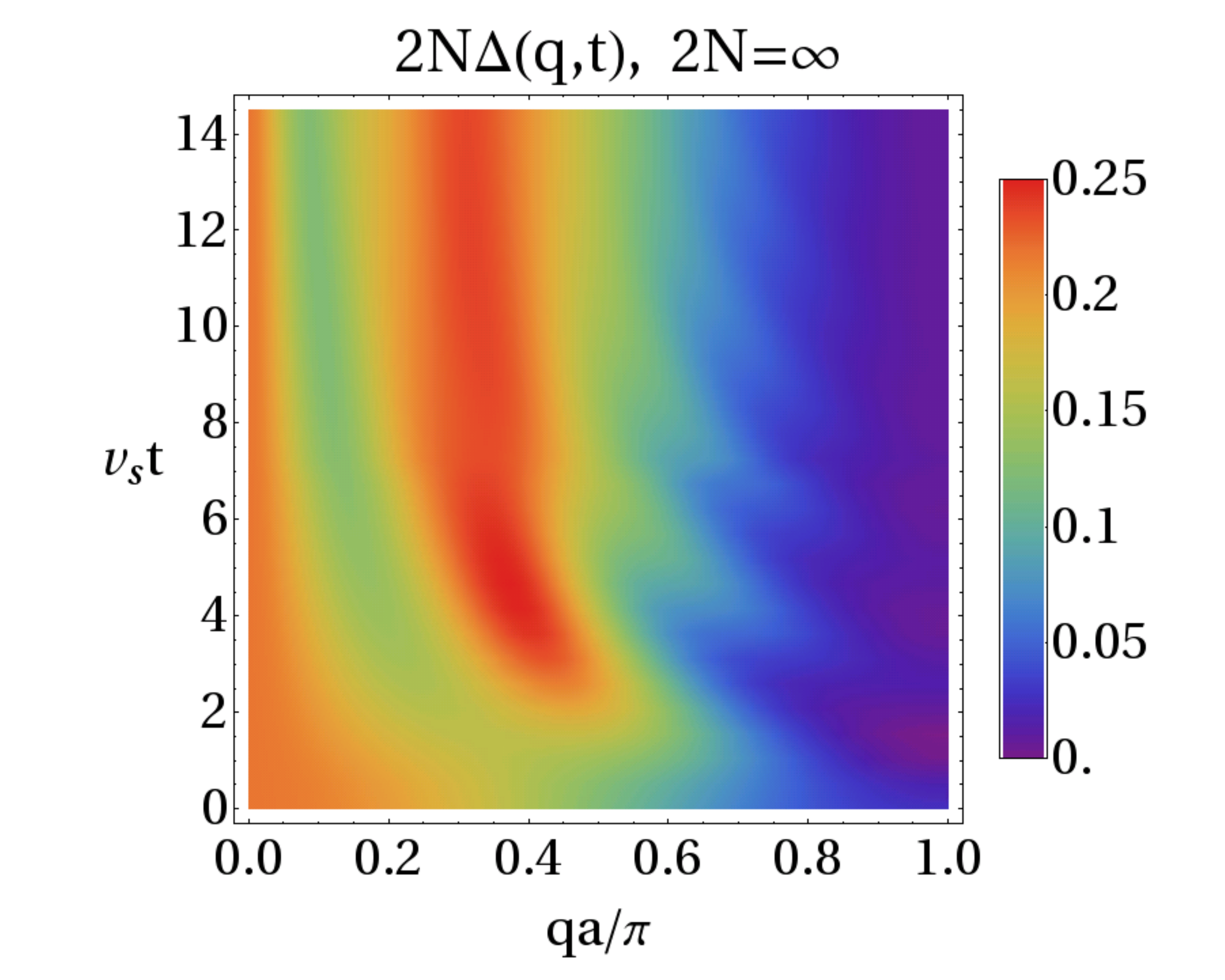}
\caption{\label{noiseplots} \figpreamble The noise correlations for the infinite lattice, TEBD simulation, $v_s=J \pi/2\hbar$.  A broad peak at $qa/\pi\sim1/2$
appears at $v_st\sim 1$, which shifts towards $qa/\pi \sim 1/3$ with the time evolution.
}
\end{center}
\end{figure}

In Fig.~\ref{noiseplots}  we plot the 'time-dependent' structure
factor $ \Delta(q,t)$, which is experimentally accessible by measuring the
noise correlations~(\ref{noise}). Alongside the persisting peak at
$q=0$, the picture  shows the formation of  a smooth peak at $q\sim
\pi/3a$ for all times $t>t^{*}$, which signals the development of
an unusual type of magnetic state. We checked that for small lattices
($2N=20$) the peak is stable for $v_st<100$. The height of the peak $N
\Delta(q,t)$ is independent of the lattice size or the type of boundary
condition, thus revealing the short-range nature of spatial
correlations in the system.

\begin{figure}[ht]
\begin{center}
\includegraphics[viewport=+140 -20 600 540,scale=0.28]{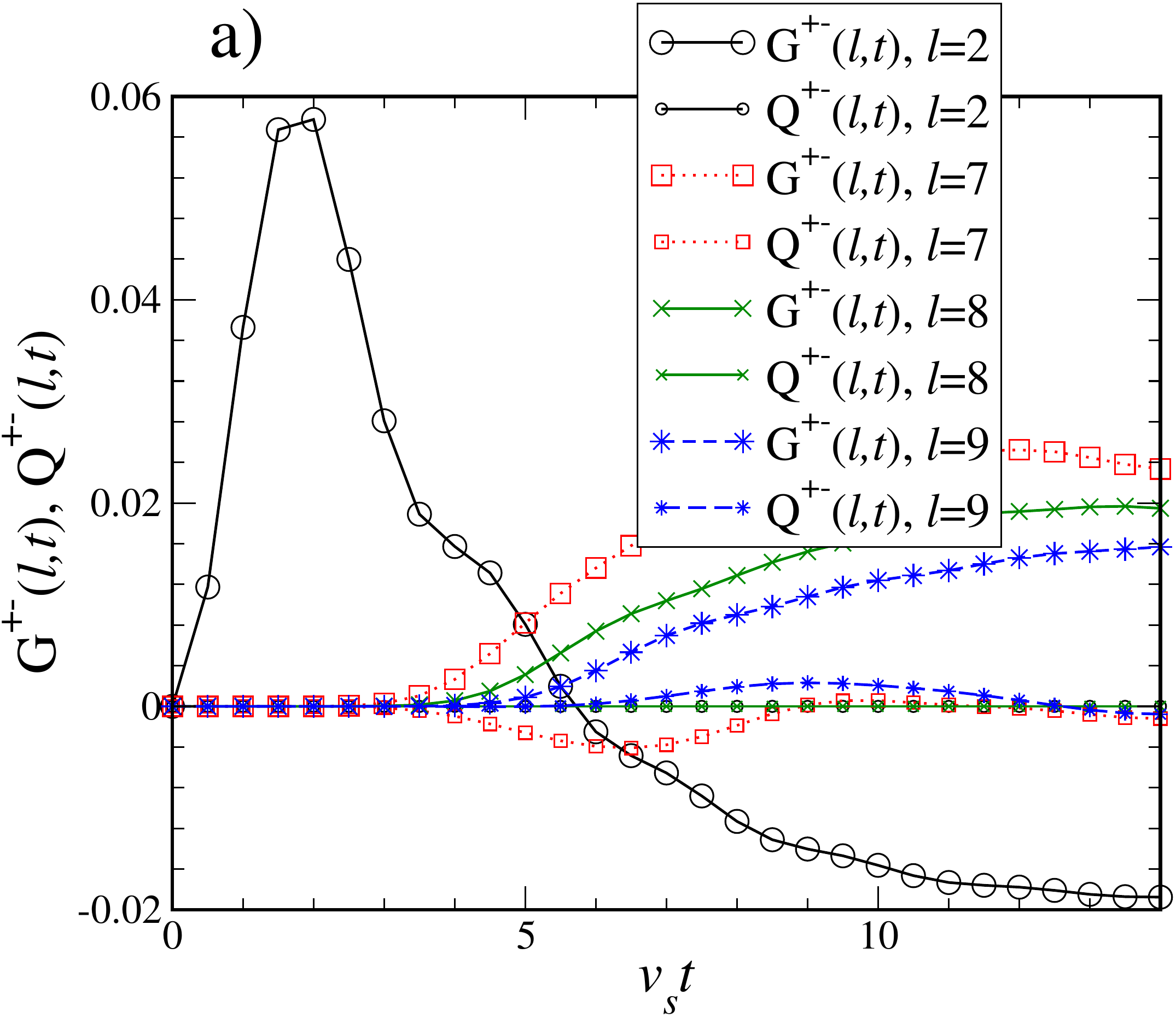}
\includegraphics[width=0.45\textwidth]{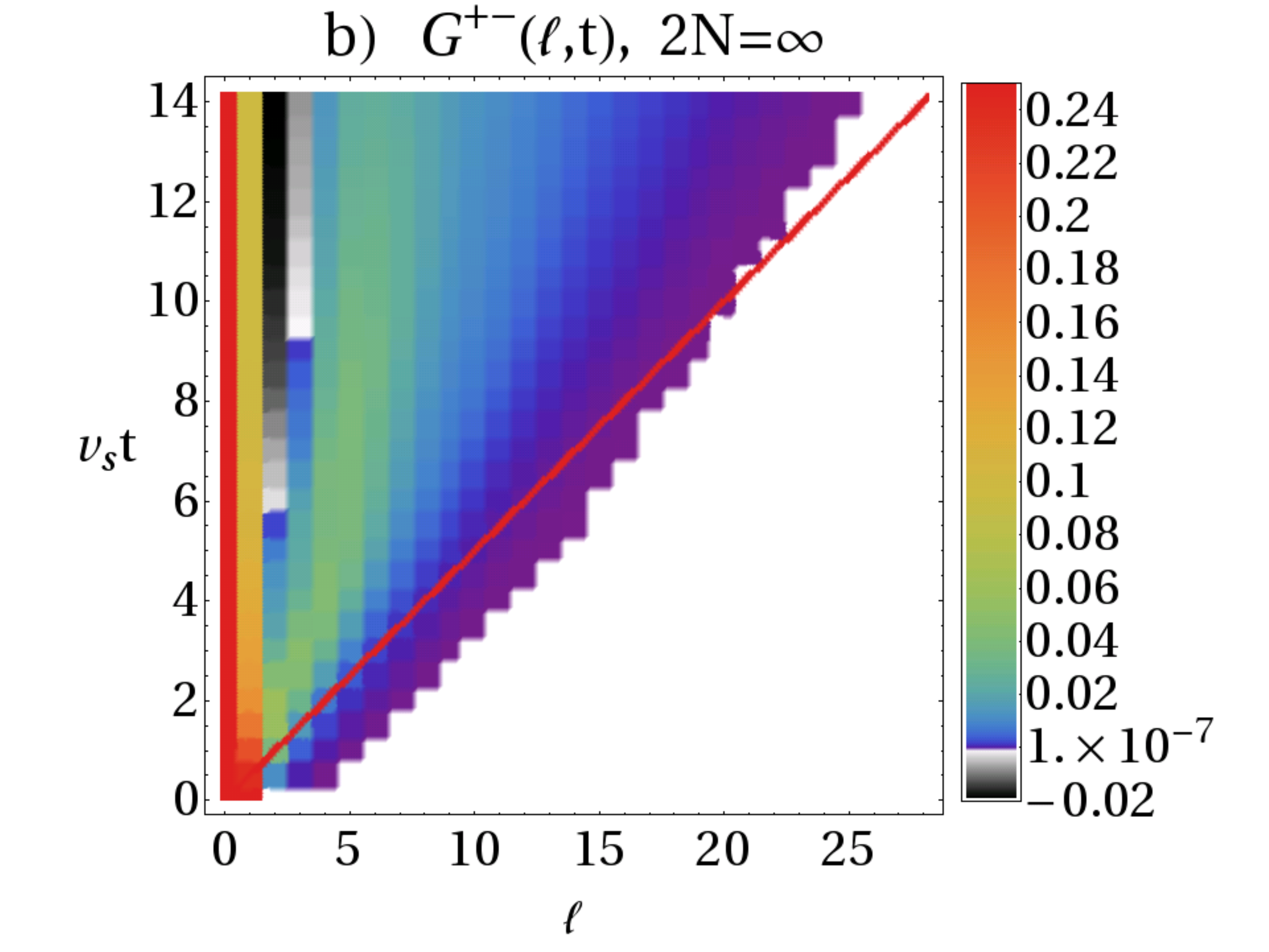}
\caption{\label{sxsx}\figpreamble Simulation of correlation functions for the
infinite lattice using TEBD, $v_s=J \pi/2\hbar$. a) Evolution of real-space
correlation functions at fixed distances: the plot shows the
tendency of longer-distance correlations to restore translational
invariance. b) Antiferromagnetic correlations at distance $l=2$,
and rapidly decaying ferromagnetic correlations for larger
distances. The straight line marks the horizon of quasiparticles moving
with spin-wave velocity $v_s$. The plot resolves magnitudes
larger than $10^{-7}$.}
\end{center}
\end{figure}

In order to explicitly study the relaxation of the correlation
functions and to understand the origin of the incommensurate peak in
the noise correlations, we plot in Fig.~\ref{sxsx} the real-space
correlation function
\beq
G^{+-}(l,t)=\sum_{|i-j|=l}\mbox{Re}\langle S^{+}_{i}(t)S^{-}_j(t)\rangle\,
\eeq
and the quantity
\beq
\label{q}
Q^{+-}(l,t)=\sum_{|i-j|=l}(-1)^i\mbox{Re}\langle
S^+_i(t)S^-_j(t)\rangle,
\eeq
which indicates that the translational symmetry is recovered for long-range
correlations. The most interesting effect we observe in the
correlation functions is the suppression of the ferromagnetic
(positive) nearest-neighbor correlations and the development of weak
antiferromagnetic (negative) correlations for next-nearest neighbor
sites. This is the origin of the incommensurate peak in the noise
correlations (Fig.~\ref{noiseplots}). The large-distance properties
of the correlation function do not contradict the predictions of
conformal field theory \cite{CC-all}. For instance, the correlations
are ferromagnetic and change from their initial values only when the
'horizon' of quasiparticle pairs $l(t)=2v_st$ passes, although
in this case we find that the horizon is not absolutely sharp. It is
important to notice that, although the horizon moves with constant speed, the intensity of the correlations decays fast with larger
distances and the correlation length remains finite.

The observed mixed correlations can be interpreted as an implication
of energy conservation. At time $t=0$ the whole correlation energy
is stored in the short-ranged triplets; at $t>0$ the action of
the evolution operator leads to the formation of longer-distance {\it
singlets} between spatially separated sites. This singlet component
persists for longer times and leads to the appearance of the
antiferromagnetic component in the spin-spin correlation function.
Therefore the local redistribution of correlation energy, revealed
in the partial AF correlation, is one possible explanation for the
emergence of mixed correlations.

\subsection{Initially prepared singlet state \label{sin}}
\begin{figure}[t]
\begin{center}
\includegraphics[width=0.45\textwidth]{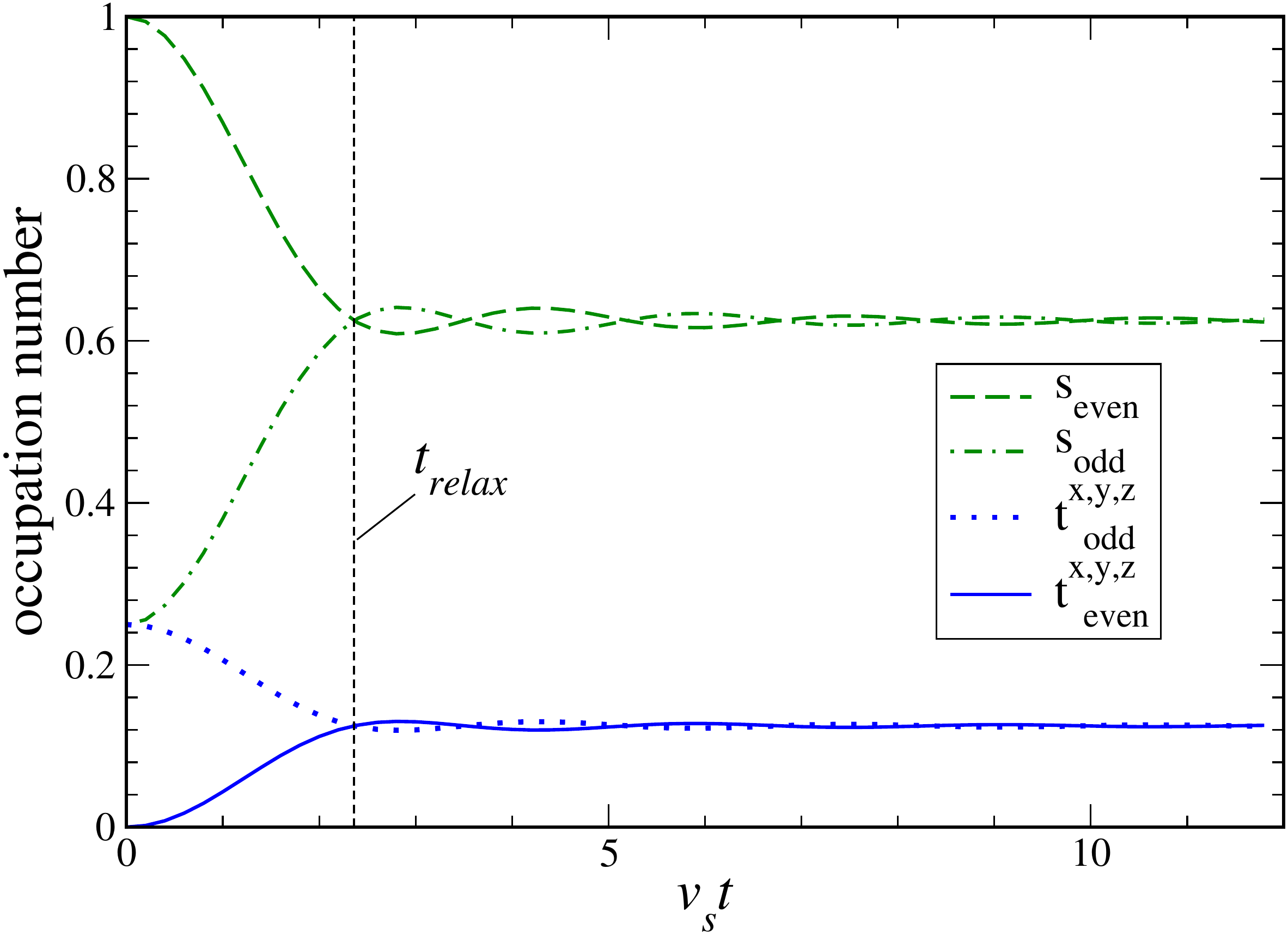}
\caption{\label{popsinglet} \figpreamble The singlet and triplet populations for
the initial singlet product state in the infinite lattice, TEBD
simulation, $v_s=J\pi/2\hbar$. The translational symmetry is recovered.}
\end{center}
\end{figure}

\begin{figure*}[t]
\begin{center}
\includegraphics[viewport=70 0 896 695,scale=0.24]{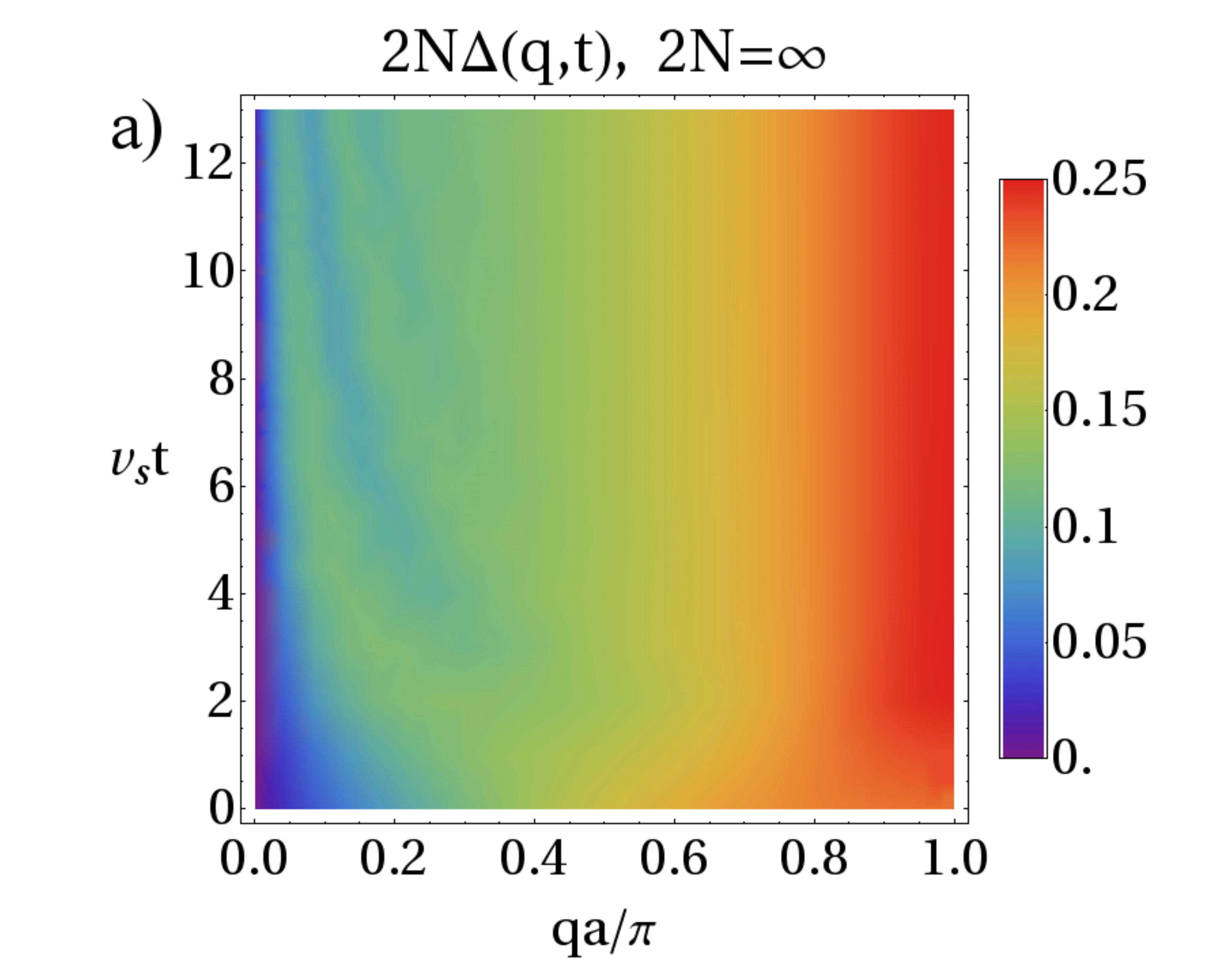}
\includegraphics[viewport=20 -10 846 630,scale=0.25]{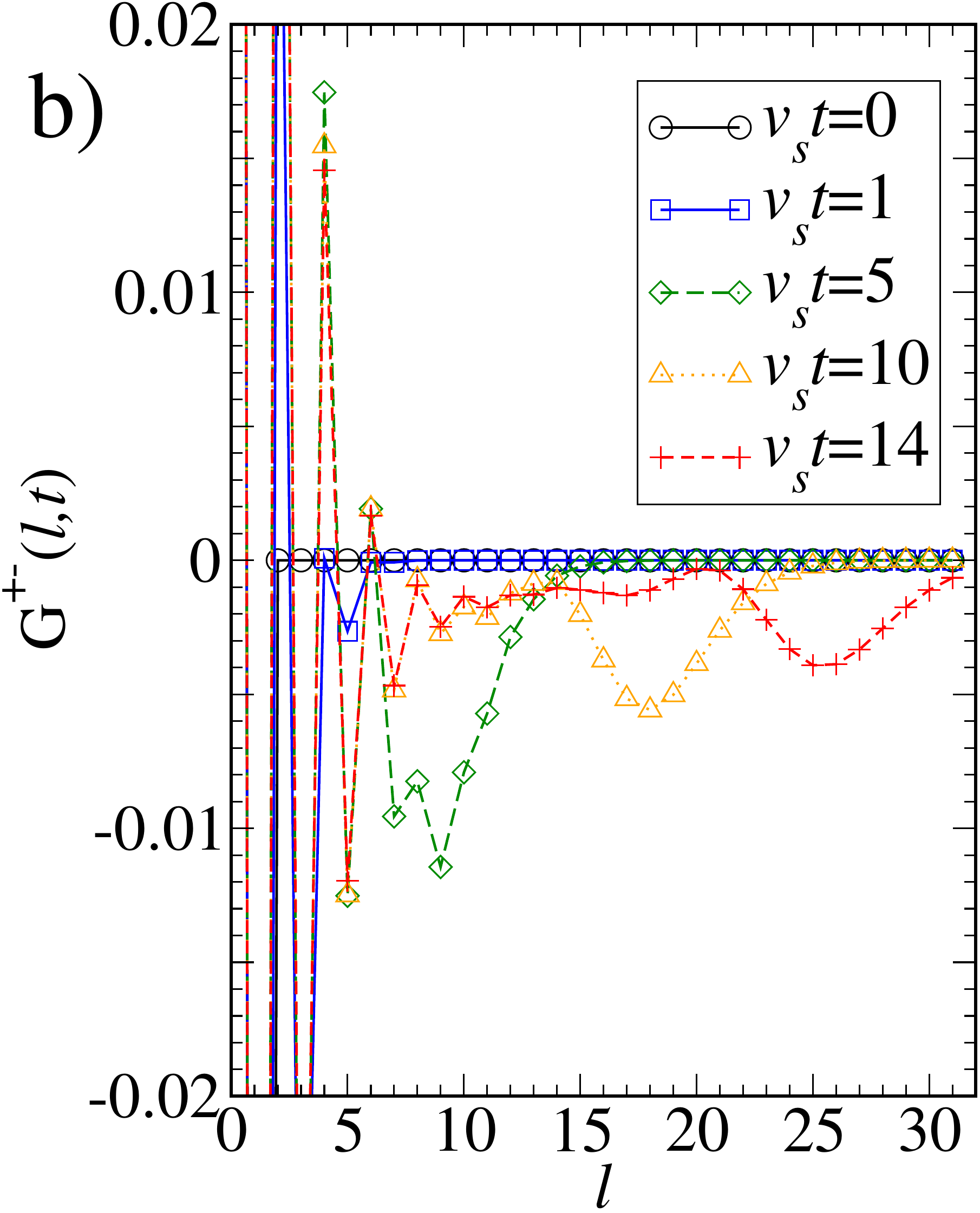}
\includegraphics[viewport=240 -15 626 600,scale=0.25]{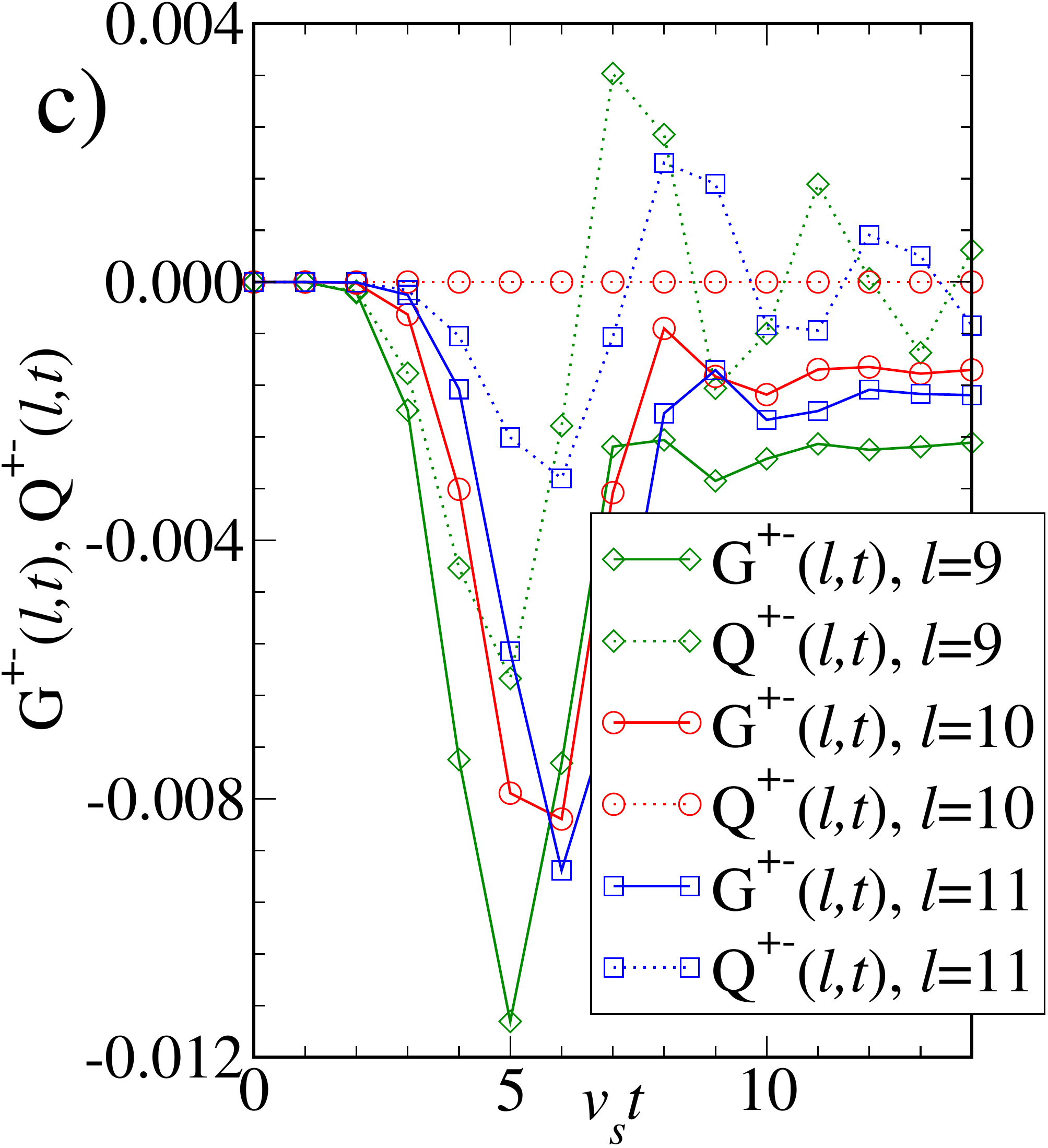}
\caption{\label{singlet}\figpreamble  Correlations for an infinite chain using TEBD simulations, $v_s=J \pi/2\hbar$.  a) The noise correlations for the system prepared in the singlet product state. Besides the
strong antiferromagetic peak there are incommensurate branches for
small $q$. b) The real-time correlation function at different moments of time.
The correlations converge to an exponentially decaying
antiferromagnetic behavior. For larger distances ($l>6$), a staggered
component centered around some finite negative value can be observed. c) The demonstration of how
the longer-range correlations remain negative after the
passing of the horizon (for even {\it and} odd distances).}
\end{center}
\end{figure*}

In this section we study  the case of the homogeneous switch
(with dynamical evolution determined by the  Heisenberg chain,
$J_1=J_2=J>0$ , Eq.~(\ref{instate})), but instead of starting from a triplet
product state,  we  now start from a $singlet$ product state,
\beq
\label{instate2}
|\psi(t=0)\rangle &=&\prod_{j}|s_{2j,2j+1}\rangle\,.
\eeq
For bosonic systems this state can be experimentally realized by
time evolution of the initial triplet product state in the presence
of a magnetic field gradient~\cite{RGBDL}. This initial state also corresponds to the decoupled double-well ground state of the
respective fermionic system, though in this case the evolving
Hamiltonian is the antiferromagnetic Heisenberg model instead of the ferromagnetic one. However, since the dynamical evolution is independent of the overall sign of the Hamiltonian, the results discussed in this
section will also hold for the fermionic system.

Unlike the case of initially prepared triplet state, here the spherical symmetry is not
broken, and the populations of the $x,y$ and $z$
components of the triplets are equal. From Fig.~\ref{popsinglet} we extract that
\beq
\label{popvalues}
&\langle t^{x,y,z}_{even}\rangle_{qs}&=\langle t^{x,y,z}_{odd}\rangle_{qs}=0.125\,,\nonumber\\
&\langle s_{even}\rangle_{qs}&=\langle s_{odd}\rangle_{qs}=0.625\,.
\eeq
These values are a direct consequence of the energy conservation,
$\frac{1}{2N}\langle H(t) \rangle =0.375$.
In Fig.~\ref{singlet} we
study the spatial correlations. Fig.~\ref{singlet}a) shows a rapidly
developed broad antiferromagnetic peak in the noise correlations and
weak incommensurate peaks at small wave vectors. These are due to
large-distance spinon correlations, depicted in
Fig.~\ref{singlet}b). The fact that the correlations remain negative
after the spinon horizon passes (Fig.~\ref{singlet}c) can be
interpreted as a memory effect of the initial singlet state.
Fig.~\ref{singlet}c) shows, by investigating the quantity $Q^{+-}(l,t)$ (see
Eq.~(\ref{q})), that the translational symmetry is recovered in the
long-range correlation functions, as is the case also for the short-range singlet and triplet correlations.

In general, the prepared singlet product state, due to its initial spherical symmetry, does not exhibit the strong mixing of anti- and ferromagnetic correlations, as the triplet state does. Although the observed
spinon correlations are interesting from the theoretical point of view, their weak effect
on the noise correlations is barely measurable experimentally. We also note that the spinon correlations may disappear on large time scales which are inaccessible numerically.

\section{Conclusion\label{concl}}

In this paper we proposed a novel protocol which creates,
from a system of two-component atoms initially prepared in an array
of triplet (singlet) pairs on neighboring sites, an array of long-distance triplet (singlet) pairs across the lattice. The method
allows parallel generation of many entangled pairs, and can
have relevant applications for the implementation of quantum
purification protocols in optical lattices. We also find that by
applying the iterative swapping procedure in an open chain one can
engineer a state in which any atom located in the right half of
the superlattice is entangled with an atom in the left half.
This state has maximally separated entangled atoms and persistency of entanglement as large as that of a cluster state, which makes it suitable for being used as a component of a one-way quantum
computer \cite{Briegel}.

We also studied the evolution of an initial triplet (singlet) product
state under a Heisenberg Hamiltonian.
Analyzing various observables we showed that while the long-range
properties of the evolving state are in agreement with those
predicted by conformal field theory, the non-universal short-range properties (e.g. the development of a magnetic state with mixed correlations), are not captured by such theoretical treatments
\cite{sengupta-2004} and have to be analyzed more carefully. They
might be a manifestation of a special type of thermalization (in the
sense of generalized Gibbs ensemble \cite{Rigol:HCBosons}), observed
in integrable systems.

The  analysis presented in this paper demonstrates that
 the coherent evolution of an initial state,
which itself can be easily  prepared -- in  our case it is just an array of
triplet (singlet) states on neighboring sites --, is a feasible
way to generate complex magnetic states with cold atoms. The
dynamical generation method is not constrained by the difficulty of
actual (physical) engineering  of exotic Hamiltonians  or by  the
low temperatures required to reach their ground states. On the other
hand, without a careful analysis it is difficult to predict a priori
the properties of the non-equilibrium state into which the system
evolves as a result of coherent quantum dynamics.
\vspace{8pt}
\section{Acknowledgements}
We would like to acknowledge Ehud Altman,  Liang Jiang, Andreas
Nunnenkamp, Oriol Romero-Isart, Michael Menteshashvili and Dionys Baeriswyl for useful discussions. This work
was partially supported by the NSF, Harvard-MIT CUA  and AFOSR. P.B.
is supported by the Swiss NSF. A.M.R. acknowledges support from
ITAMP.

\end{document}